\documentclass[%
 reprint,
superscriptaddress,
 amsmath,amssymb,
 aip,
longbibliography
]{revtex4-2}
\usepackage[utf8]{inputenc}
\usepackage{graphicx}
\usepackage{epstopdf}
\usepackage{amsmath,amssymb}
\usepackage{xcolor}
\usepackage{graphicx}
\usepackage{dcolumn}
\usepackage{bm}
\usepackage{hyperref}

\begin{document}

\preprint{AIP/123-QED}

\title{
Early warnings of tipping in a non-autonomous turbulent reactive flow system: efficacy, reliability, and warning times\\
}

\author{Ankan Banerjee}
\email{ankan1090@gmail.com}
\affiliation{Department of Aerospace Engineering, Indian Institute of Technology Madras, Chennai 600036, India}  
\author{Induja Pavithran}%
\email{indujap2013@gmail.com}
\affiliation{Department of Aerospace Engineering, Indian Institute of Technology Madras, Chennai 600036, India}
\author{R. I. Sujith}%
\email{sujith@iitm.ac.in}
\affiliation{Department of Aerospace Engineering, Indian Institute of Technology Madras, Chennai 600036, India}%

\date{\today}

\begin{abstract}
Real-world complex systems such as the earth's climate, ecosystems, stock markets, and combustion engines are prone to dynamical transitions from one state to another, with catastrophic consequences. State variables of such systems often exhibit aperiodic fluctuations, either chaotic or stochastic in nature. Often, the parameters describing a system vary with time, showing time dependency. Constrained by these effects, it becomes difficult to be warned of an impending critical transition, as such effects contaminate the precursory signals of the transition. Therefore, a need for efficient and reliable early-warning signals (EWS) in such complex systems is in pressing demand. Motivated by this fact, in the present work, we analyze various EWS in the context of a non-autonomous turbulent thermoacoustic system. In particular, we investigate the efficacy of different EWS in forecasting the onset of thermoacoustic instability (TAI) and their reliability with respect to the rate of change of the control parameter. We consider the Reynolds number ($Re$) as the control parameter, which is varied linearly with time at finite rates. The considered EWS are derived from critical slowing down, spectral properties, and fractal characteristics of the system variables. The state of TAI is associated with large amplitude acoustic pressure oscillations that could lead thermoacoustic systems to break down. We consider acoustic pressure fluctuations as a potential system variable to perform the analysis. Our analysis shows that irrespective of the rate of variation of the control parameter, the Hurst exponent and variance of autocorrelation coefficients warn of an impending transition well in advance and are more reliable than other EWS measures. Additionally, we show the variation of the warning time to an impending TAI with rates of change of the control parameter. We also investigate the variation of amplitudes of the most significant modes of acoustic pressure oscillations with the Hurst exponent. Such variations lead to scaling laws which could be significant in prediction and devising control actions to mitigate TAI.
\end{abstract}

\maketitle

\begin{quotation}
The ability of early warning signals (EWS) to provide true warnings for an impending, yet unwanted, transition from one state to another in real-world complex systems still remains an outstanding problem, especially when a system is non-autonomous, \textit{i.e.}, the control parameter is time-dependent, and the dynamics of the system have inherent fluctuations. This work attempts to build an understanding of the efficacy of different early warning signals (EWS) and their robustness in providing true alarms in the context of a non-autonomous turbulent thermoacoustic system prone to thermoacoustic instability (TAI). The phenomenon of TAI has catastrophic consequences as it could lead thermoacoustic systems such as gas turbine engines used for power production, liquid rocket engines, or jet engines used for aviation to break down. The emergence of TAI is due to positive feedback between the unsteady heat release rate fluctuations and the acoustic pressure. Such turbulent systems have high levels of inherent fluctuations. We use EWS derived from critical slowing down, spectral properties, and fractal characteristics of the thermoacoustic system. According to our analysis, when a complex system such as a turbulent thermoacoustic system is non-autonomous and whose dynamics are permeated with fluctuations, fractal-based and spectral-based EWS are more robust and reliable in providing warnings than the classical critical slowing down-based early warning signals.
\end{quotation}


\section{Introduction}
Real-world complex systems, natural or human-made, often experience shifts from one state to another when the system crosses a critical threshold in one of its components (which could either be an external condition governing the system or the system variables)~\cite{vanNes:TEE:2016}. Such a phenomenon is known as tipping~\cite{Gladwell_tipping_book:2006}. Positive feedback between such components propels accelerating change in the dynamics of the underlying system towards an alternative state once the critical threshold is passed. Among numerous natural systems, the Earth's climate~\cite{dakos_PNAS:2008}, and various ecosystems~\cite{scheffer_Nature:2001,pedersen_RSOS:2017} have experienced several abrupt transitions in the past. Apart from natural systems, transitions in the economic and industrial systems, such as the collapse of stock markets~\cite{may_Nature:2008}, and breakdown in power grids or blackouts~\cite{ren:EPSR:2015,suchithra:Chaos2020} are of immense concern. The phenomenon of tipping has been observed even in physiological systems, namely, cardiac arrest~\cite{Ivanov_Nature:1999}, epileptic seizures~\cite{Mcsharry_NatMed:2003}, and cancer metastasis~\cite{sarkar_PNAS:2019}. Occasionally, the impact of tipping is catastrophic and, in the worst case, irreversible. Therefore, it is crucial to be alert to the possibility of impending tipping.

Ashwin \textit{et al.}~\cite{ashwin_PhilTracRSA:2012} have classified tipping into three categories - bifurcation-induced tipping (B-tipping), noise-induced tipping (N-tipping) or rate-induced tipping (R-tipping). In the case of B-tipping, the change of states of a complex system is realized by means of the evolution of a system variable with respect to the external condition (which are attributed as parameters in mathematical models) driving the system. The parameter value at which a transition occurs is known as a tipping point~\cite{scheffer_NATURE:2009}. The occurrence of tipping can either be continuous or abrupt and is defined as supercritical and subcritical transitions, respectively, in the bifurcation theory~\cite{strogatz_nonlinear_book:2018}. Mathematically, at the tipping point, the real part of an eigenvalue of the underlying stable state of a dynamical system crosses the imaginary axis from the left. Since the real part of an eigenvalue approaches zero in the close vicinity of a tipping point, the return time of the system to a small perturbation becomes larger. This phenomenon, which is called `critical slowing down (CSD)~\cite{scheffer_Nature:2001}, plays a pivotal role in indicating upcoming transitions in a wide class of systems. Generic features of CSD have been captured by well-known statistics such as lag-1 autocorrelation, recovery rate and variance~\cite{scheffer_NATURE:2009}, skewness~\cite{guttal_EL:2008}, and kurtosis~\cite{Biggs_PNAS:2009}. These statistics serve the purpose of potential early warning signals (EWS) to a tipping point. Additionally, spectral measures~\cite{bury_RSI:2020,dakos_PloSone:2012,pavithran_SciRep:2020} possessing signatures of CSD also serve as early warning indicators.

Apart from B-tipping, the presence of random fluctuations in a system due to noise and the rapid change of another parameter (known as the rate-sensitive parameter) may lead a system to tip to an alternate state without altering the driving or bifurcation parameter, causing N-tipping and R-tipping respectively. Unlike the case of B-tipping, it is challenging to get warnings based on classical CSD-based indicators in the other two classes of tipping. Tipping in real-world complex systems can be due to any combination of these three types of tipping phenomena. As a result, identifying tipping points and their precursory signals in such systems is really challenging. Identifying robust EWS and how reliable they are in providing warnings of an impending tipping phenomenon is still an open area of research.

Another constraint in getting EWS is the availability of data. Occasionally, the time series extracted from components of events of open systems, such as the climate system or ecological systems, is sparse and not well sampled. Even the sample size and window size have effects on determining EWS. Closed environment (or isolated) experiments having highly sampled data, where one has control over the parameters of the system, could be a platform for developing robust indicators. We propose turbulent reactive flow systems such as thermoacoustic systems as potential complex systems in resolving the issues of sparse data, sampling size, repeatability, etc., arising in various real-world systems. Such systems are amenable to efficient laboratory experiments and could generate highly sampled data. Apart from that, studying the transitions in thermoacoustic systems is highly relevant and has immediate applications. 

Combustors, as they happen to be thermoacoustic systems, are indispensable components of power and aviation industries~\cite{lieuwen2021unsteady}. Such systems are complex, having nonlinear interactions between the acoustic field of the combustor, the underlying hydrodynamic field, and the flame~\cite{sujith_TI:2021}. Due to the positive feedback between the acoustic pressure and the heat release rate fluctuations in thermoacoustic systems, thermoacoustic instability (TAI) emerges in the system in the form of large amplitude oscillations. 

Initially, the transition to TAI occurs while changing system parameters such as the heater power, airflow rate, or fuel flow rate. The occurrence of TAI has been reported by varying the control parameter in a static/quasi-static manner. Interestingly, N-tipping and R-tipping to TAI have also been observed in different thermoacoustic systems~\cite{jegadeesan_ntipping:2013,manikandan_afterburner:2020}. The  stable operating state of a turbulent combustor exhibits low amplitude aperiodic fluctuations as opposed to the steady quiescent state in the case of laminar systems. The stable operating state is known as `combustion noise' in the thermoacoustic community. Nair and \textit{et al.}~\cite{nair_IJSCD:2013}, and Tony \textit{et al.}~\cite{tony_PRE:2015} showed that the state of combustion noise in turbulent thermoacoustic systems is associated with a high-dimensional chaotic attractor. 
 
Traditionally, researchers focus on the quasi-static variation of the control parameter for studying different characteristics of TAI. Also, engineers establish stability margins in quasi-static experiments of a thermoacoustic system. However, the control parameter in most thermoacoustic systems is changed continuously in real-life situations. The continuous variation of the control parameter, instead of a quasi-static variation, makes the thermoacoustic system a non-autonomous one. Interestingly, analyzing the characteristics of non-autonomous thermoacoustic systems becomes possible as these systems are amenable to rigorous laboratory experiments. 
 
Tony \textit{et al.}~\cite{tony_SciRep:2017} showed that the onset of TAI could be advanced depending on the initial configuration of the underlying thermoacoustic system and the rate of change of the control parameter. 
Bonciolini \textit{et al.}~\cite{bonciolini_RSOS:2018} observed rate-induced tipping delay in a turbulent combustion system for faster rates of change of the control parameter. The authors also found that the system has a transition from the state of combustion noise to a state of thermoacoustic instability via a subcritical Hopf bifurcation. Due to the presence of turbulent fluctuations in each state of the system, the authors repeated an experiment one hundred times to compute the probability distribution of the amplitude of oscillations for identifying rate-induced delay characteristics. Although the delay in the transition to TAI is induced due to the ramping of the control parameter, Unni \textit{et al.}~\cite{unni_Chaos:2019} have shown that in the case of slower or intermediate variation of the control parameter, the initial condition and noise have nontrivial effects on such characteristics. Depending on the noise level, a system may undergo N-tipping. For the same rate of change of parameter, the transition to TAI occurs at different parameter values for different realizations of an experiment.

Although there is a plethora of research on tipping phenomena and their precursors in thermoacoustic systems, literature on the classical EWS indicators of an impending TAI is still limited~\cite{gopalakrishnan_SciRep:2016,An_JEGTP:2019,pavithran_review_EPJST:2021}. Gopalakrishnan \textit{et al.}~\cite{gopalakrishnan_SciRep:2016} used CSD-based indicators in providing warnings to TAI in a laminar thermoacoustic system. Recently, Pavithran and Sujith~\cite{pavithran2021effect} have shown the performance of different EWS in providing precursors to TAI as well as warning times with respect to the rate of change of control parameter in a non-autonomous laminar thermoacoustic system. They reported that the warning time decreases, and the warning in parameter increases with an increase in the rate of change of the control parameter. However, the performance and reliability of different EWS indicators in practical turbulent thermoacoustic systems remain unanswered. 

Since one has control over governing parameters, a systematic investigation of turbulent thermoacoustic systems approaching TAI could provide a qualitative understanding of the dynamics of various real-world complex systems approaching tipping. Therefore, analyzing the functionality of different EWS in thermoacoustic systems could help to comprehend their strength in providing warnings in those open systems. The present work, in this context, analyzes the efficacy and reliability of different EWS with respect to the rates of change of the control parameter.
 
First, we identify the onset of TAI from the time series of acoustic pressure fluctuations obtained at different ramping rates from experiments on the turbulent combustor. Subsequently, we analyze the performance of EWS based on critical slowing down, fractal characteristics, and spectral properties of the time series of acoustic pressure fluctuations. The reliability of different EWS is then tested using the receiver operating characteristics (ROC) at different rates. Further, we identify different scaling laws which can help extrapolate numerous system characteristics at the tipping point.

\section{Experiments}
\begin{figure}[t]
    \centering
    \includegraphics[height=!,width=0.48\textwidth]{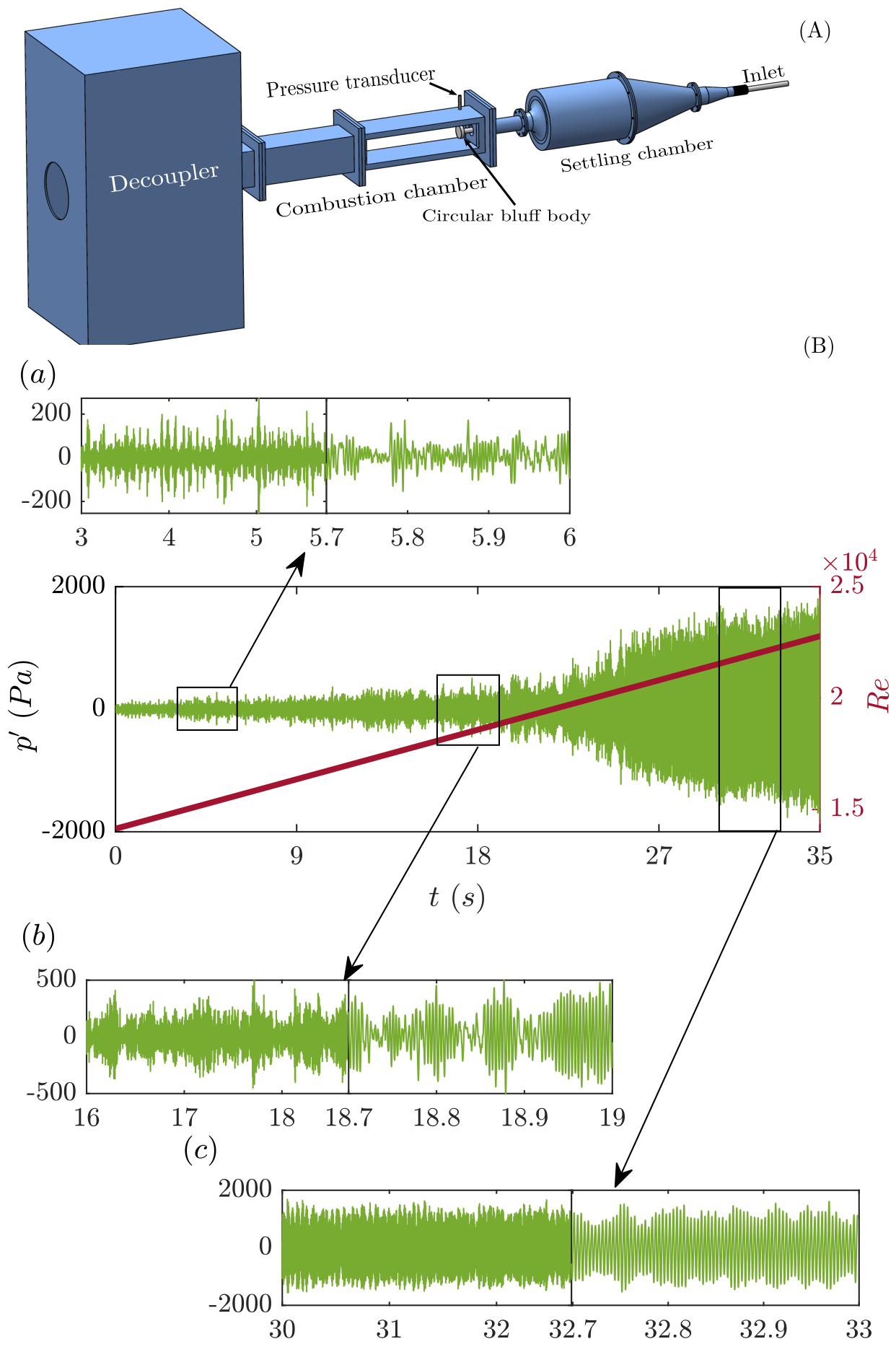}
    \caption{ (A) Schematic of the turbulent combustor used in the current study. Acoustic pressure fluctuations are measured using a piezoelectric pressure transducer. The design of the combustor was adapted from Komarek and Polifke~\cite{komarek_combustor_design:2010}. (B) The light green curve represents a prototypical time series of acoustic pressure fluctuations ($p^\prime$), and the variation of the Reynolds number ($Re$) is shown in brown. States of (a) combustion noise, (b) intermittency, and (c) thermoacoustic instability are observed, respectively, as the control parameter Reynolds number is increased.}
    \label{fig:schematic_diagram}
\end{figure}

In the present work, we perform experiments in a thermoacoustic system which is a partially premixed turbulent combustor. The schematic of the combustor is shown in Fig.~\ref{fig:schematic_diagram}A. The combustor consists of several components such as an inlet, a settling (plenum) chamber, a combustion chamber (a horizontal duct with one end open) having a circular disk mounted on a central shaft, also known as the bluff-body, and a decoupler at the open end of the duct. The decoupler is used to reduce the radiation of acoustic energy from the combustor to the surroundings. 
The combustion chamber has a length of $700$ mm with a square cross-section of $90 \times 90~\text{mm}^2$. The bluff body has a radius of $47~\text{mm}$ and a thickness of $10$ mm.

A piezoelectric sensor (PCB103B02, sensitivity: $217.5$ mV/kPa, resolution: $0.2$ Pa and uncertainty: $0.15$ Pa) flush-mounted on the wall of the combustor near the bluff body measures the acoustic pressure fluctuations in volts (V) at a sampling rate of $12~\text{kHz}$. Mass flow controllers (Alicat Scientific, MCR Series) are used to measure and control mass flow rates of air $(\Dot{m_a})$ and fuel $(\Dot{m_f})$ with an uncertainty of $\pm(0.8\%$ of the reading $+0.2\%$ of the full scale). The Reynolds number (Re) for the reactive flow is obtained as $Re = 4 \Dot{m} /\pi \mu (D_0+D_1),$ where $\Dot{m}=\Dot{m_a}+\Dot{m_f}$ is the mass flow rate of the air-flow mixture, $(D_0)$ is the diameter of the burner, $(D_1)$ is the diameter of the circular bluff-body and $\mu$ is the dynamic viscosity of the air-fuel mixture in the experimental conditions. Reynolds number for the reported experiments is considered as the control parameter and varied within a range from $Re=(1.41 \pm 0.08) \times  10^4$ to $Re=(3.0  \pm 0.26) \times 10^4$. More details of the experimental setup are discussed in work by Raghunathan \textit{et al.}~\cite{raghunathan_JFM:2020}.

Since the focus here is on the study of non-autonomous thermoacoustic systems, we consider the control parameter $Re$ as a function of time $t$ such that $Re=Re_0+rt$ where $Re_0$ is the initial value of Reynolds number and $r=d Re/d t$ denotes the rate of change of control parameter. Initially, near $Re_0$, the combustor is at the state of combustion noise where acoustic pressure fluctuations $(p^\prime)$ show low-amplitude aperiodic oscillations. Thermoacoustic instability appears in the system at higher values of $Re$. During the state of TAI, acoustic pressure fluctuations $p^\prime$ show high amplitude periodic oscillations. We observe a transition from the state of combustion noise to the state of TAI via intermittency in a single realization of an experiment while increasing the Reynolds number ($Re$) monotonically from $Re_0$. The state of intermittency is characterized by bursts of periodic oscillations appearing apparently randomly amongst epochs of low-amplitude aperiodic oscillations~\cite{nair_thampi_JFM:2014}. A prototypical time series of an experiment and its different states (light green curves) are plotted in figure~\ref{fig:schematic_diagram}(B, a-c) together with the variation of $Re$ (the brown-colored curve). 

\section{Results}
\subsection{Identifying the onset of thermoacoustic instability:}
In thermoacoustic systems, the state of thermoacoustic instability is associated with stable limit cycle attractors. 
It has been reported that a stable fixed point in a laminar thermoacoustic system, the horizontal Rijke tube, becomes unstable to a finite amplitude limit cycle via a subcritical Hopf bifurcation~\cite{subramanian_IJSCD:2010,gopalakrishnan_SciRep:2016}.
Authors in those works accomplished their investigation by varying the control parameter, the heater power, either at every $20$ s~\cite{gopalakrishnan_SciRep:2016} or in a quasi-static manner~\cite{subramanian_IJSCD:2010}. Later, Etikyala and Sujith~\cite{etikyala_Chaos:2017} showed that the transition from a stable fixed point to a limit cycle could have a dependence on a secondary parameter of the system, which was the airflow rate in their case. Depending on the airflow rate supplied to the Rijke tube, the transition can occur via either a supercritical or a subcritical Hopf bifurcation. 

These preceding works are of the kind of bifurcation-induced tipping. Tipping points were the critical heater power at which the system bifurcates from the quiescent state to thermoacoustic instability. On the contrary, it is unlikely that we can detect a state transition by identifying the bifurcation point when the control parameter is changed continuously, i.e., when the system is non-autonomous. In such cases, the system does not get enough time to converge to the stable attractor associated with the underlying autonomous dynamical system. As a result, a continuous state transition is observed with the variation of the parameter~\cite{unni_Chaos:2019,pavithran2021effect,bonciolini_NonDyn:2019,bonciolini_RSOS:2018}. Moreover, trajectories either approach or hover around the stable attractors of the underlying autonomous dynamical system within their respective basin of attractions for a finite amount of time. In order to understand the characteristics of different early warning signals and their proximity in providing subsequent control actions, first, we identify the onset of thermoacoustic instability in the case of a non-autonomous thermoacoustic system.

In a non-autonomous system, the occurrence of a state transition or tipping can be determined by identifying the unstable attractor forming a basin boundary between the two attractors of the underlying autonomous system. The parameter value at which a trajectory crosses that basin boundary would be considered the tipping point. 
In a non-autonomous laminar thermoacoustic system, Tony \textit{et al.}~\cite{tony_SciRep:2017} showed that tipping could occur at different parameter values, depending on the initial conditions and the rate of change of the control parameter, which they referred to as as preconditioned R-tipping. Further, there exists a critical rate above which the onset of transition is advanced in comparison to the tipping point from the underlying autonomous system.  In real-world non-autonomous systems such as thermoacoustic or aeroacoustic systems, it is strenuous to determine the basin boundary from the time series of measurable system variables. As an alternative method to identify tipping, Pavithran and Sujith~\cite{pavithran2021effect} studied a non-autonomous horizontal Rijke tube and investigated tipping to TAI by analyzing the rate of change of the root-mean-square value of the acoustic pressure fluctuations (${p^\prime}_{rms}$). They found that the rate of change of ${p^\prime}_{rms}$ showed a discontinuity when thermoacoustic instability occurs. The critical heat power at which such discontinuity occurs serves as the tipping point in their investigation.

To determine the tipping point in the present scenario, we first identify the critical Reynolds number ($Re_c$) at which the system moves into the basin of attraction of TAI. Then, we define $Re_c$ as the onset of thermoacoustic instability. Any trajectory of a system variable with some initial condition will approach stable limit cycle attractors corresponding to the underlying autonomous system once it moves past the critical Reynolds number $Re_c$. However, the trajectory cannot settle to such attractors as we vary $Re$ linearly with time $t$ at a finite rate $r$. Instead, it would hover around the attractors in a helical path which can be realized in three dimensional $Re$-velocity bifurcation diagram.     
Additionally, the variable would move with $Re$ by forming a helical trajectory. By analyzing the time series of the rate of change of ${p^\prime}_{rms}$, we are unable to determine $Re_c$ due to the presence of inherent fluctuations in the system. Earlier works have reported the presence of low frequencies in the amplitude spectrum of the acoustic pressure fluctuations during the state of combustion noise~\cite{nair_JFM:2014}. Moreover, the fundamental acoustic mode of the combustor becomes dominant during the state of TAI~\cite{pavithran_SciRep:2020}. These results motivate us to analyze the spectral properties of the acoustic pressure signal for determining $Re_c$. Toward this, we perform continuous wavelet transform of the time series of acoustic pressure fluctuations.
\begin{figure}[ht]
    \centering    
    \includegraphics[height=!,width=0.49\textwidth]{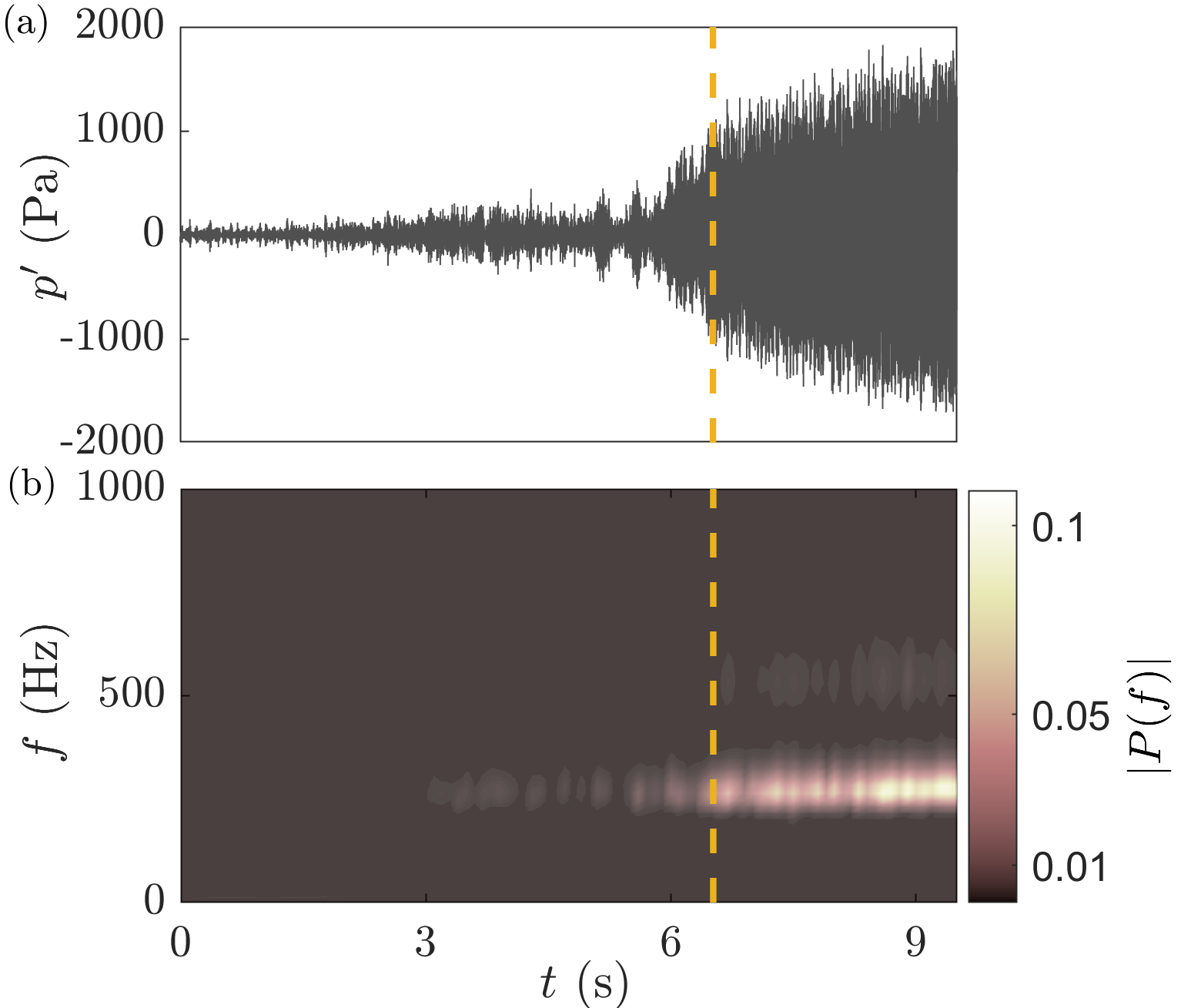}
    \caption{(a) Time series of acoustic pressure fluctuations for ramping rate $r=2283~\text{s}^{-1}$. (b) Power spectrum of frequencies corresponding to the time series in (a). The graph shows that frequencies $250-300~\text{Hz}$ become dominant during the state of thermoacoustic instability. Initially, the frequency band $250-300~\text{Hz}$ appears intermittently and becomes continuous once the state of TAI is achieved. The color bar represents the power of frequencies.}
    \label{fig:wavelet_freq}
\end{figure}

We use the \textit{Morse} wavelet~\cite{lilly_Morse_wavelets:2008} as the mother wavelet for the transformation and the standard MATLAB package \textit{cwt} to evaluate the wavelet transform. Figure~\ref{fig:wavelet_freq} shows the spectrogram of a representative time series of the  acoustic pressure fluctuations. At the beginning of an experiment, i.e., when the pressure signal is aperiodic (corresponding to the state of combustion noise~\cite{nair_JFM:2014}), we observe that the power spectrum is broadband, as evident from the initial dark gray region. The natural frequency $(f)$ of the combustor duct is approximately $275~\text{Hz}$. The power spectrum of the frequency band $250-300~\text{Hz}$ starts to become significant at intermittent intervals as the Reynolds number is increased, which is evident from lighter spots within the dark region in figure~\ref{fig:wavelet_freq}. The spectrum within the range of $250-300~\text{Hz}$ in the spectrogram becomes brighter, representing strong oscillations, as $Re$ is increased further. Finally, the spectrum forms a continuous band as the Reynolds number crosses a critical value (corresponding to the time marked by the dashed yellow line). This represents that the system moves into the state of periodic oscillations associated with TAI. Therefore, the Reynolds number at the time stamp ($t_c$), marked by the yellow dashed line, is considered to be the critical parameter value ($Re_c$) where a state transition occurs. The peak spectral amplitude becomes sharper and narrower, which is evident from the light yellow and red color region in figure~\ref{fig:wavelet_freq}.
 
\subsection{Effects of the ramping rate $(r)$ on $Re_c$:}
After fixing the critical Reynolds number $(Re_c)$ at which TAI onsets, we analyze the effects of the rate of change of control parameters and turbulent fluctuations on the onset. According to earlier studies, ramping the control parameter and/or noise has non-trivial effects on the tipping point. Jegadeeshan and Sujith~\cite{jegadeesan_ntipping:2013} showed a noise-induced transition from stable to oscillatory state in a ducted non-premixed flame. Depending on the strength of the noise, the transition occurred earlier than the bifurcation-induced transition. On the other hand, ramping of the control parameter delays the onset of TAI~\cite{tony_SciRep:2017,bonciolini_RSOS:2018,unni_Chaos:2019} due to the effect of inertia. As a result, a trajectory starting close to an initial stable state remains adjacent to it even for parameter values greater than the tipping point of the underlying autonomous system. However, the effects of inertia become insignificant in the presence of noise in the system~\cite{unni_Chaos:2019}. Since noise incorporates stochasticity in the dynamics, the system can randomly tip to the alternative state.

Interestingly, the effect of the inertia can also be attenuated in a deterministic system, i.e., in the absence of noise, if the underlying stable state is a high-dimensional chaos in nature. The projection of a high-dimensional chaotic state is reflected on a system variable as fluctuations that seem like random fluctuations. Nair \textit{et al.}~\cite{nair_IJSCD:2013} have shown that the aperiodic state of the time series representing combustion noise is high dimensional chaos. Later, Tony \textit{et al.}~\cite{tony_PRE:2015} confirmed the deterministic nature of the combustion noise, which is different from the measurement (white) as well as dynamic (colored) noises present in the system. Based on these results, we can say that the time series of acoustic pressure fluctuations in a turbulent combustor is deterministic in nature and permeated with measurement and dynamic noises.
\begin{figure}[ht]
    \centering    
    \includegraphics[height=!,width=0.49\textwidth]{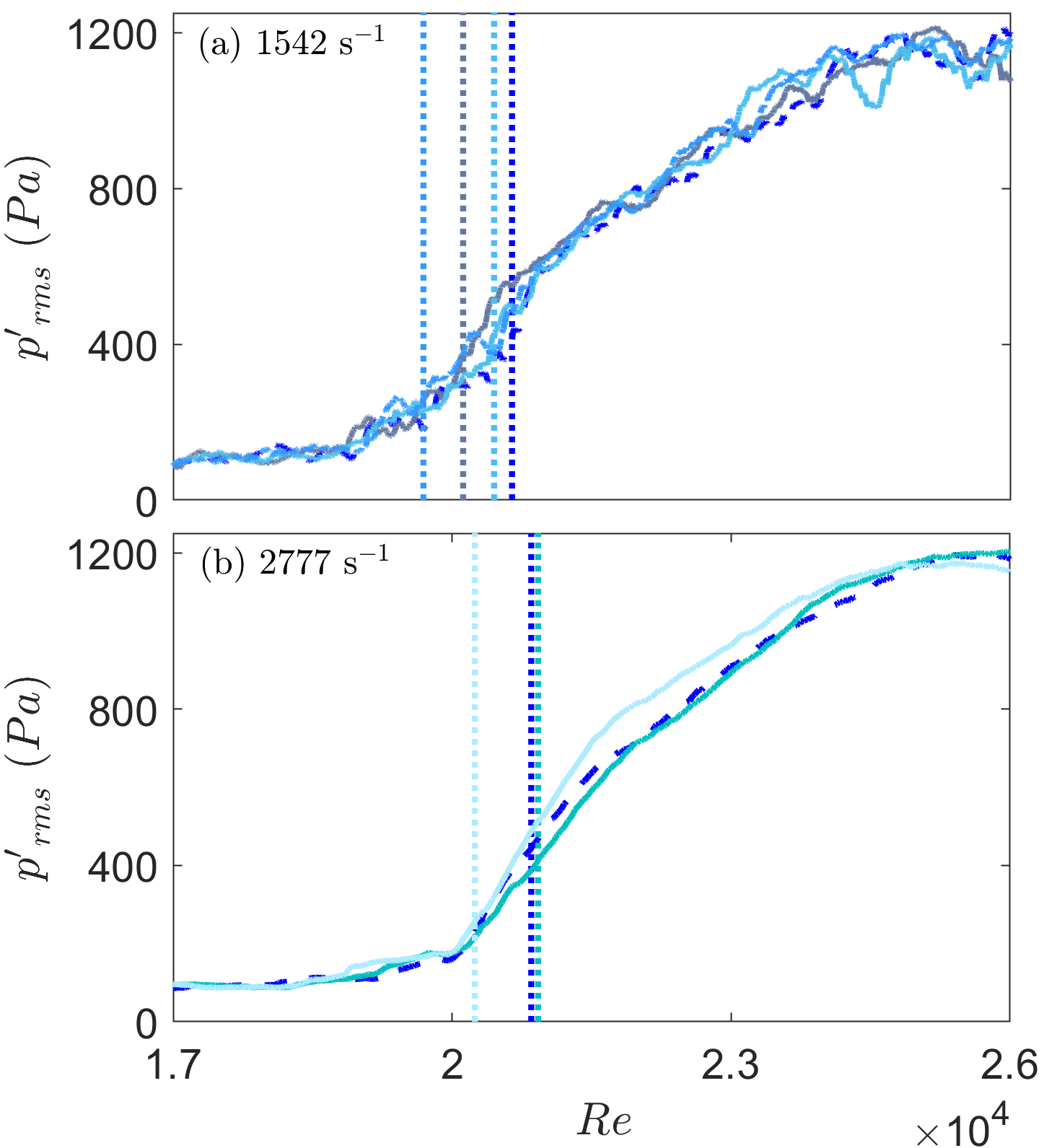}
    \caption{Plots of ${p^\prime}_{rms}$ of acoustic pressure fluctuations in the combustor as obtained from several realizations of the same experiment (shown in different colors) corresponding to the ramping rate (a) $r=1542~\text{s}^{-1}$ and (b) $r=2777~\text{s}^{-1}$. The onset of TAI (dashed vertical lines) differs for experiments with the same values of $r$ due to turbulence.}
    \label{fig:prms_re_1060}
\end{figure}
 Therefore, we can expect that the onset of the thermoacoustic instability $Re_c$ will vary for different independent realizations, which have a fixed value of the ramping rate $r$. This implies that $Re_c$ will not show any monotonic behavior even if $r$ is changed monotonically. A similar kind of observation has been reported earlier by Unni \textit{et al.}~\cite{unni_Chaos:2019} in the case of a laminar thermoacoustic system, the horizontal Rijke tube.
 
\begin{figure}[ht]
    \centering
    \includegraphics[height=!,width=0.48\textwidth]{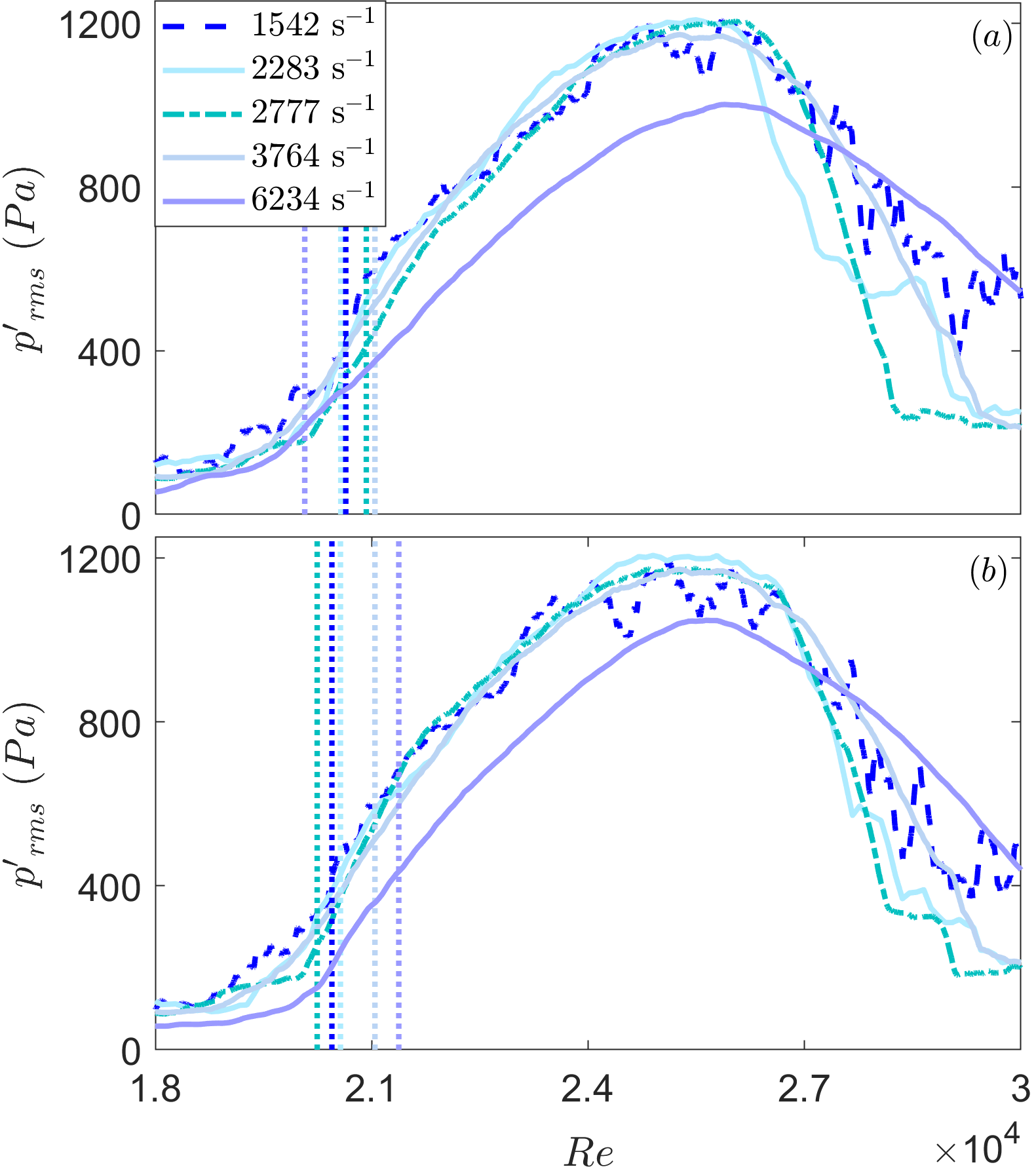}
    \caption{ Comparison between the onset of TAI for different values of $r$ is shown. 
    Dashed, dot-dashed, and solid lines in different shades of blue represent the time series of ${p^\prime}_{rms}$ for the ramping rates $r=1542,~2283,~2777,~3764,~\text{and}~6234~\text{s}^{-1}$. (a) and (b) signify random non-monotonic behavior of $Re_c$ (dotted lines) for different realizations of the same set of experiments. The variation in $Re_c$ is due to turbulence.}
    \label{fig:prms_re_rates}
\end{figure}

 To highlight those facts, we plot the time series of ${p^\prime}_{rms}$ for (a) $r=1542~\text{s}^{-1}$ and (b) $r=2777~\text{s}^{-1}$ in figure~\ref{fig:prms_re_1060} for different realizations of the same experiment. We observe that the onset of the thermoacoustic instability for different experiments indicated by dashed vertical lines varies for a fixed value of $r$. Subsequently, we plot several realizations of the time series of ${p^\prime}_{rms}$ for different values of the ramping rate $r$. From figure~\ref{fig:prms_re_rates}(a) and~\ref{fig:prms_re_rates}(b), it is evident that $Re_c$ does not change monotonically for a monotonic variation in $r$ which is due to the presence of turbulent fluctuations in the system. However, the average behavior (ensemble average) of a turbulent system may still show the rate-delayed characteristics for higher values of $r$~\cite{bonciolini_RSOS:2018}.
 Since our focus is on the behavior of different EWS, we will be dealing with individual realizations of an experiment for the rest of the study.

\subsection{Performance of different EWS: \label{EWS}}
Early warning signals analyzed in the present analysis are variance ($VAR$), skewness ($SKEW$), kurtosis ($K$) (\textit{i.e.} moments of order 2, 3, 4, respectively), autocorrelation for lag-1 ($AC$), the variance of autocorrelation ($var(AC)$), the Hurst exponent ($H$), and the moment of order two ($\mu_2$). To compute these EWS from the time series, we consider a moving window of size $0.6$ seconds and slide it at $0.1$ seconds for lower rates and $0.01$ seconds for higher rates. The selection of the window size is detailed in the Supplementary Material. Although it is challenging to identify CSD from empirical evidence, several statistical quantifiers serve the purpose well. Among them, autocorrelation at different lags, variance, skewness, and kurtosis have shown convenient signatures in identifying CSD in numerous climatic and ecological regime shifts~\cite{dakos_PloSone:2012}. Theoretical~\cite{scheffer_NATURE:2009} and empirical~\cite{dakos_PNAS:2008} evidence show that, due to CSD, both autocorrelation and variance increase when a system approaches a tipping point. In addition to autocorrelation and variance, skewness and kurtosis can signify CSD and are used as EWS~\cite{guttal_EL:2008}. A nonzero skewness represents asymmetry in the distribution, which happens when a system leans towards an alternative state. Higher skewness suggests that the system is closer to the tipping point. Kurtosis quantifies the thickness of the tail of a distribution associated with a time series. Moreover, spectral properties of the time series data also contain signatures of CSD~\cite{dakos_PloSone:2012,pavithran_SciRep:2020}.  
 Apart from CSD-based indicators, spectral and fractal characteristics of a system variable are worth an investigation to get a warning of an impending transition. Fractal properties of the human heartbeat are used to classify a healthy heart state from a diseased one~\cite{Ivanov_Nature:1999}. It is also used to quantify volatility in stock markets~\cite{calvet_multifractality:2002}. Recently, Pavithran et al.~\cite{pavithran_review_EPJST:2021} have shown that moments of power spectra, the variance of autocorrelations $(var(AC))$ of a finite number of lags,  and the Hurst exponent $(H)$ can be used as early warning indicators to an impending thermoacoustic instability.

\begin{figure*}[t]
    \centering\includegraphics[height=0.85\textwidth,width=0.8\textwidth]{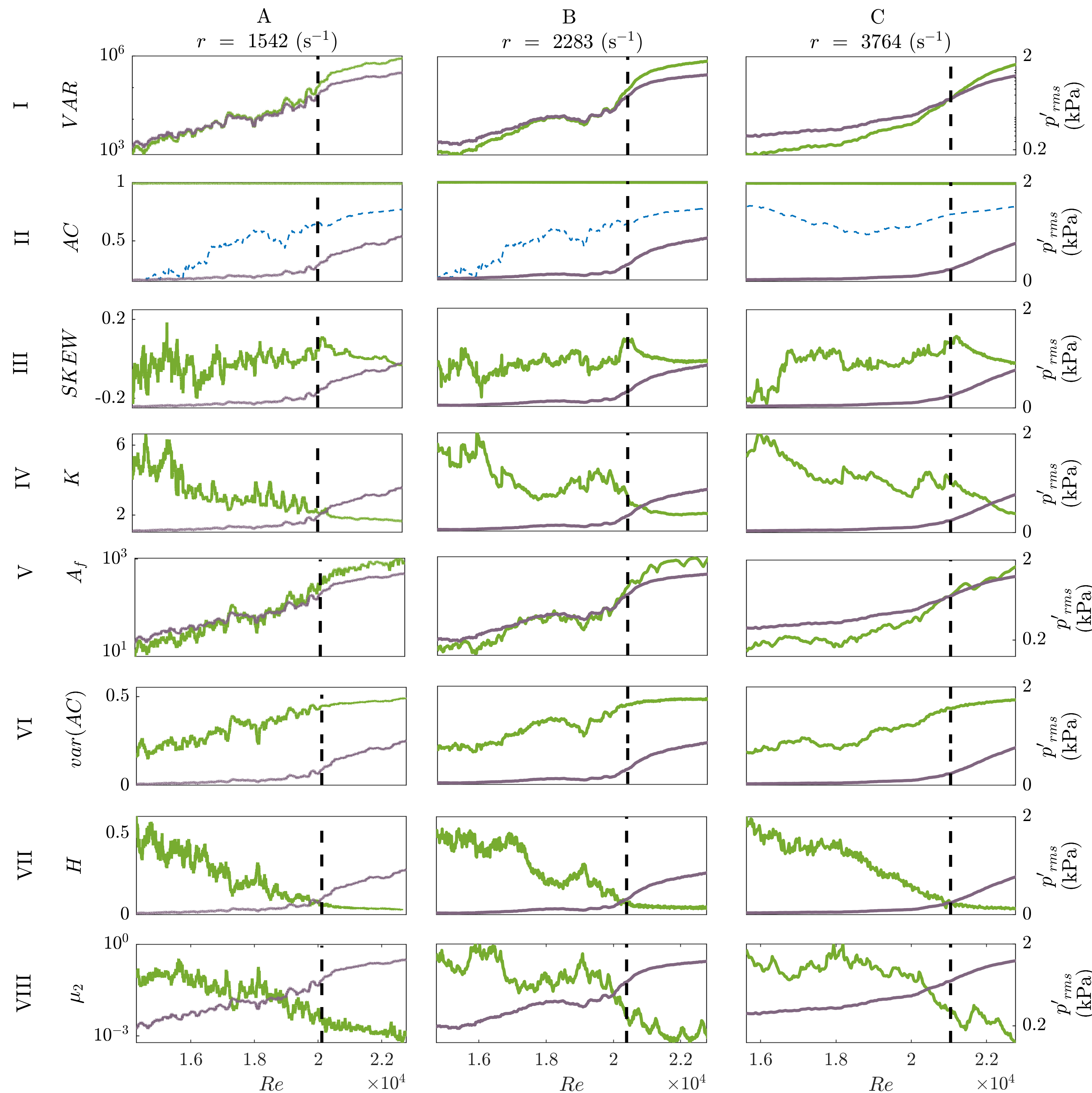}
    \caption{The figure reveals the proximity in providing warnings to TAI for different EWS together with the effects of the rate of change of the control parameter ($r$) on those measures. Different EWS are shown in light green colored curves for three ramping rates, (A) $r=1542~\text{s}^{-1}$, (B) $r=2283~\text{s}^{-1}$, and (C) $r=3764\text{s}^{-1}$. The EWS are (I) variance ($VAR$), (II) lag-1 and lag-44(dashed blue curves) autocorrelation ($AC$), (III) skewness ($SKEW$), (IV) kurtosis ($K$), (VI) variance of autocorrelations of different lags ($var(AC)$), (VII) Hurst exponent ($H$), and (VIII) spectral moment of order two ($\mu_2$). The ${p^\prime}_{rms}$ for different rates are shown in violet. Black dashed lines represent the onset of TAI. (V) The amplitude of the dominant spectral frequency ($A_f$) is shown in light green.  An increase in $VAR$, signifying an approaching transition, is observed for different rates. A weak variation in $AC$ is observed for higher rates (II-C); otherwise, it remains constant. $SKEW$ shows inconsistency in providing EWS. $K$ signals a transition weakly for higher rates (IV-B) and (IV-C). $var(AC),~H,$ perform convincingly for different values of $r$ as EWS as evident from VI and VII. The area under the ROC curve (AOC) is also calculated for different measures at different rates.
    }
    \label{fig:10-40-100-slpmpers}
\end{figure*}

 In figure~\ref{fig:10-40-100-slpmpers}, we have shown these different EWS for three different ramping rates, namely, $1542, ~2283,$ and $3764~\text{s}^{-1}$. Black dashed lines represent the onset of thermoacoustic instability ($Re_c$). Curves in violet correspond to ${p^\prime}_{rms}$ of the acoustic pressure fluctuations ($p^\prime$). From figure~\ref{fig:10-40-100-slpmpers}(I), it follows that $VAR$ can detect the onset of instability (${Re}_c$) in advance as it starts to grow prior to the onset of instability. The rise in skewness, indicating an impending transition, is evident from figure~\ref{fig:10-40-100-slpmpers}(III). However, the characteristic is prominent for higher rates only. The variation in kurtosis with $Re$, on the other hand, is unable to indicate an impending transition, as evident from figure~\ref{fig:10-40-100-slpmpers}(IV).
 
  We observe from figure~\ref{fig:10-40-100-slpmpers}(II) that $AC$ remains almost constant, with values close to $1$ for all ramping rates we investigate. The constant behavior or the weak variation of $AC$ discerns that the time series of acoustic pressure fluctuations $(p^\prime)$ is highly correlated irrespective of the system's state under investigation. This is possibly due to the presence of turbulent fluctuations in the system. Interestingly, the autocorrelation of the lag of one acoustic cycle (\textit{i. e.} lag-44 in our case) increases inconsistently for lower rates and is very close to the onset of TAI for higher rates. Thus identifying significant shortest time scales, perhaps, could improve the performances of autocorrelation as a warning signal.

Although autocorrelations for different lags show no/weak signals to tipping, their statistics could forewarn the tipping point efficiently. Bury \textit{et al.}~\cite{bury_RSI:2020} showed spectral characteristics of autocorrelations for higher lags that can serve as an EWS. Recently, Pavithran \textit{et al.}~\cite{pavithran_review_EPJST:2021} proposed the variance of autocorrelations for several lags $(var(AC))$ could be an effective EWS to TAI in a turbulent combustor. They showed that $var(AC)$ grows gradually as the system approaches the onset of TAI. Since the state of TAI is associated with periodic oscillations, the periodic nature is also reflected in the autocorrelation coefficient ($\rho$) for several lags. Moreover, the maximum variation in $\rho$ is observed during a periodic oscillation, in which case $var(AC)$ attains its maximum value of 0.5. In a similar manner, here, we observe (figure~\ref{fig:10-40-100-slpmpers}(VI)) that $var(AC)$ shows an increment in its value as the system approaches a critical transition to the state of TAI. Unlike in the case of $p_{rms}$, the change in $var(AC)$ prior to TAI occurs well in advance and persists for different ramping rates we have considered. Although $var(AC)$ provides warnings to TAI robustly for different ramping rates, its computation is constrained by the selection of an appropriate number of lags. Therefore, considering a large number of lags may generate redundancy into $var(AC)$. To overcome such situations, we tested with different values of $\tau$ and found that lags $(\tau)$ between two to four acoustic cycles served our purpose well. The selection of $\tau$ is detailed in the Supplementary Material. 

The Hurst exponent $(H)$ has a prognostic value in providing a warning to an impending TAI~\cite{nair_JFM:2014}. As the system proceeds towards TAI from the state of combustion noise, the periodic content in the time series increases. As a result, the value of $H$ decreases toward zero. Therefore, a decrease in the value of $H$ is considered a warning of an impending TAI. The Hurst exponent $H$ has been found to be a robust EWS for all ramping rates assumed in the experiment with a non-autonomous horizontal Rijke tube~\cite{pavithran2021effect}. In the present analysis, we also observe the same characteristic of $H$ as the control parameter $Re$ approaches $Re_c$. It is evident from figure~\ref{fig:10-40-100-slpmpers}(VII) that for ramping rates $r=1542,~2283,$ and $3764~\text{s}^{-1}$, $H$ (solid violet curves) decreases as the onset of TAI (black dashed line) is approached. Moreover, the decrement begins much earlier than the occurrence of TAI. Such characteristics prevail for a wide range of ramping rates that we have investigated. Thus, the Hurst exponent $H$ is an effective EWS measure for forecasting the onset of TAI in non-autonomous turbulent thermoacoustic systems. In this context, we can infer that $H$ can work efficiently as an EWS for a transition from chaos to order or vice-versa in any non-autonomous complex systems with fluctuations that are deterministic in nature.

Among the various spectral-based EWS, the spectral measure has been proven to be a reliable EWS in thermoacoustic, aeroacoustic, and aeroelastic systems~\cite{pavithran_SciRep:2020}. The values of the spectral moment of order $j~(\mu_j)$ decrease (for one or several values for $j$) as the system approaches a limit cycle. Here, we found that the spectral moment of second order (${\mu}_2$) is a potential EWS and is shown in figure~\ref{fig:10-40-100-slpmpers}(VIII).

So far in the discussion, we have presented a comparison between different EWS. It is evident from figure~\ref{fig:10-40-100-slpmpers} that only a few of the investigated EWS can actually provide warnings, irrespective of rates, to avert the transition. Among the conventional EWS, $AC,\text{ and}~K$ could not provide warnings of an impending TAI, irrespective of rates. Variance and skewness provide warnings for different $r$. However, skewness shows a weak increase prior to tipping and is not robust with respect to the ramping rate $r$. On the other hand, non-conventional measures such as $H, var(AC),\text{ and},~\mu_2$, as well as $VAR$, are robust to provide warnings for different values of $r$.  

After determining the robustness of different EWS, we look into the reliability of these measures for different values of the rate of change of the control parameter. In this regard, we compute the receiver operating characteristics (ROC) curves for the time series of EWS from figure~\ref{fig:10-40-100-slpmpers} and calculate the area under the ROC curve (AUC). The next section details the derivation of ROC curves and AUC.

\subsection{Receiver Operating Characteristics (ROC):}
Receiver operating characteristics measure the true positive rate or sensitivity of a binary classifier. Curves corresponding to ROC, first developed in signal-processing literature~\cite{keller2009bioeconomics,green1966signal}, represent the false positive rate at any detected true positive rate. Such curves are a two-dimensional quantifier of the reliability of a given binary classifier. A similar one-dimensional quantifier is obtained by measuring the area under the ROC curve (AUC).

\begin{figure*}[th]
    \centering
    \includegraphics[height=!,width=0.8\textwidth]{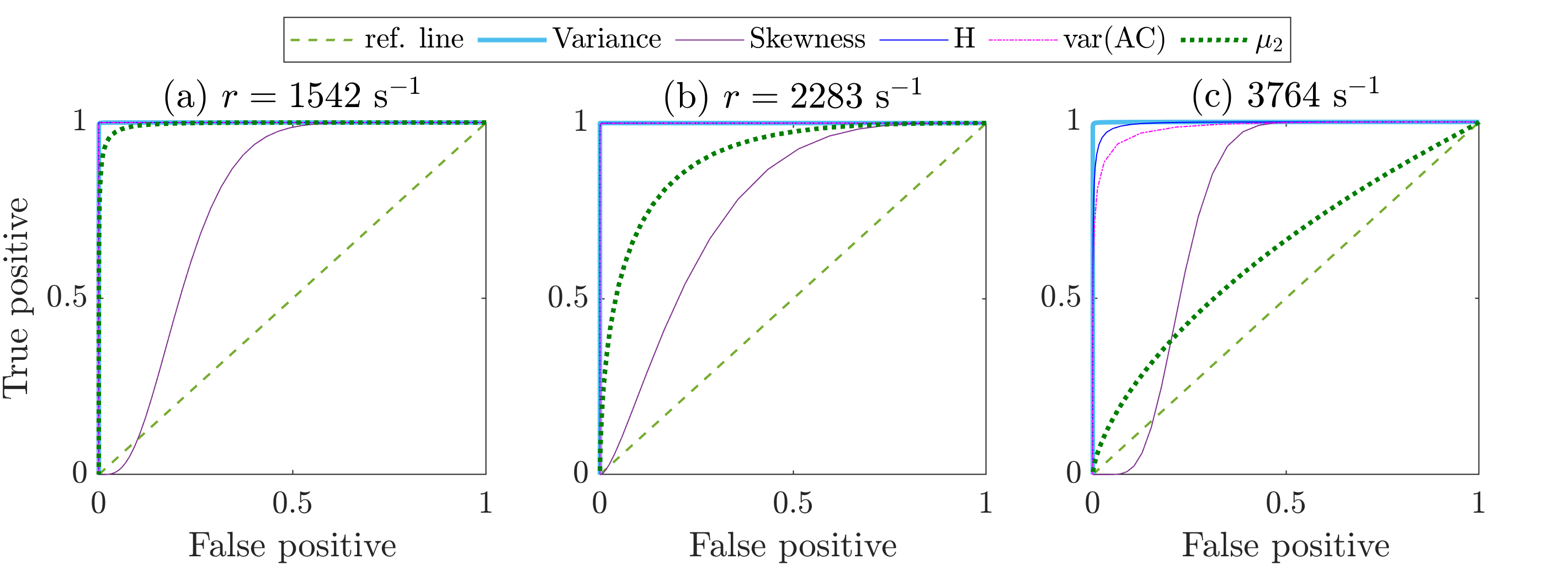}
    \caption{ROC curves for different EWS, namely the Hurst exponent (yellow solid curve), the variance of autocorrelations $(\text{blue dotted curve})$, variance $(\text{orange dashed curve})$, spectral moments of order two $(\text{pink solid curve})$, skewness (blue solid curve), kurtosis (brown dot-dashed curve), and lag-1 autocorrelation (violet dotted curve) are shown at different values of $r$, (a) $1542~\text{s}^{-1}$, (b) $2283~\text{s}^{-1}$, and (c) $3764~\text{s}^{-1}$. Concavity of curves justifying their reliability as EWS.}
    \label{fig:ROC_clc}
\end{figure*}
In the context of tipping, various EWS can be considered as a binary classifier separating the stable state from a state approaching tipping. The reliability of an EWS depends on how well it can classify these two states. True (false) positive rates are calculated at each indicator value serving as a potential threshold for an impending transition. For EWS indicators, a ROC curve is constructed based on the distribution of an indicator at the stable state (attributed as the null state in the literature) and at a state approaching a transition (attributed as the test state in the literature)~\cite{boettiger_RSIf:2012}. Generally, the distribution curves overlap, and false positive rates signify the overlapped region. If the distribution curves overlap exactly with each other, then the sensitivity and the false positive rate have the same value, and the ROC curve becomes a straight line. The more the overlap region, the more severe the trade-off between true and false positive rates.

In the present analysis, the state of combustion noise is considered a stable or null state, and the state of intermittency close to the onset is considered a test state. We determine distributions of null and test states from the same time series. The initial goodness of an indicator depends on how better it can sense a test state for all threshold values. Moreover, as we are changing the control parameter $Re$ continuously, each threshold value of an indicator has a time stamp. As a result, each threshold value corresponds to a warning time which is the time interval between the time stamp of the threshold value and the time stamp of the onset of TAI. Therefore, ROC gives the trade-off between the true positive rate and the false positive rate of warning times given by underlying indicators such as $H,~p_{rms},~var(AC)$ in the present context. ROC acts as a quantifier to measure false positivity in an EWS.  

The ROC curves for ramping rates $1542,~2283,~\text{and}~3764~\text{s}^{-1}$ are shown in  figure~\ref{fig:ROC_clc} (a), (b), and (c), respectively. The violet and blue-colored thin curves, the cyan-colored thick curve, the magenta-colored dashed-dotted curve, the green dotted thick curve, and the green dashed thin curve represent ROC curves corresponding to $SKEW,~H,~VAR,~var(AC),~\text{and}~\mu_2$, and the line of reference, respectively. From figure~\ref{fig:ROC_clc}, we observe that the ROC curves are concave down for all indicators, except for skewness, at different values of $r$. It is evident from figure~\ref{fig:ROC_clc} that reliable EWS are $VAR,~var(AC)$, $H$, and $\mu_2$ since true positive rates are always higher than false positive rates. All curves, except $\mu_2$, and $SKEW$, are more concave and almost similar to each other. Values of skewness crosses the reference line, which implies that it could provide false warnings. It is evident from figure~\ref{fig:ROC_clc} that the concavity of ROC curves for $VAR,~var(AC)$, and $H$ are higher than $\mu_2$. Therefore, they are considered to be more reliable indicators of an upcoming TAI. Although the warning time given by these different indicators varies, and even some of the indicators can provide very low warning time to take control actions, ROC curves show the reliability of tested indicators at different values of $r$. 

We also compute AUC values corresponding to ROC curves. The following table shows computed AUC values for significant EWS for different values of $r$:

\begin{center}
\begin{tabular}{ c c c c }
 EWS & $r=1542$ & $r=2283$
  & $r=3764$ \\
  & (s$^{-1}$) & (s$^{-1}$) & (s$^{-1}$)\\
 \hline
 $VAR$ & 0.92 & 0.99 & 0.98 \\  
 $SKEW$ & 0.76 & 0.78 & 0.77\\
 $H$ & 0.99 & 0.99 & 0.99 \\
 $\mu_2$ & 0.9 & 0.92 & 0.9
\end{tabular}
\label{Table_AUC}
\end{center}
Note here that although skewness has a higher value of AUC, it is not a reliable measure since its ROC curves show the possibility of providing false alarms.

\subsection{Warning time to thermoacoustic instability:}
\begin{figure}[h]
    \centering
    \includegraphics[height=!,width=0.48\textwidth]{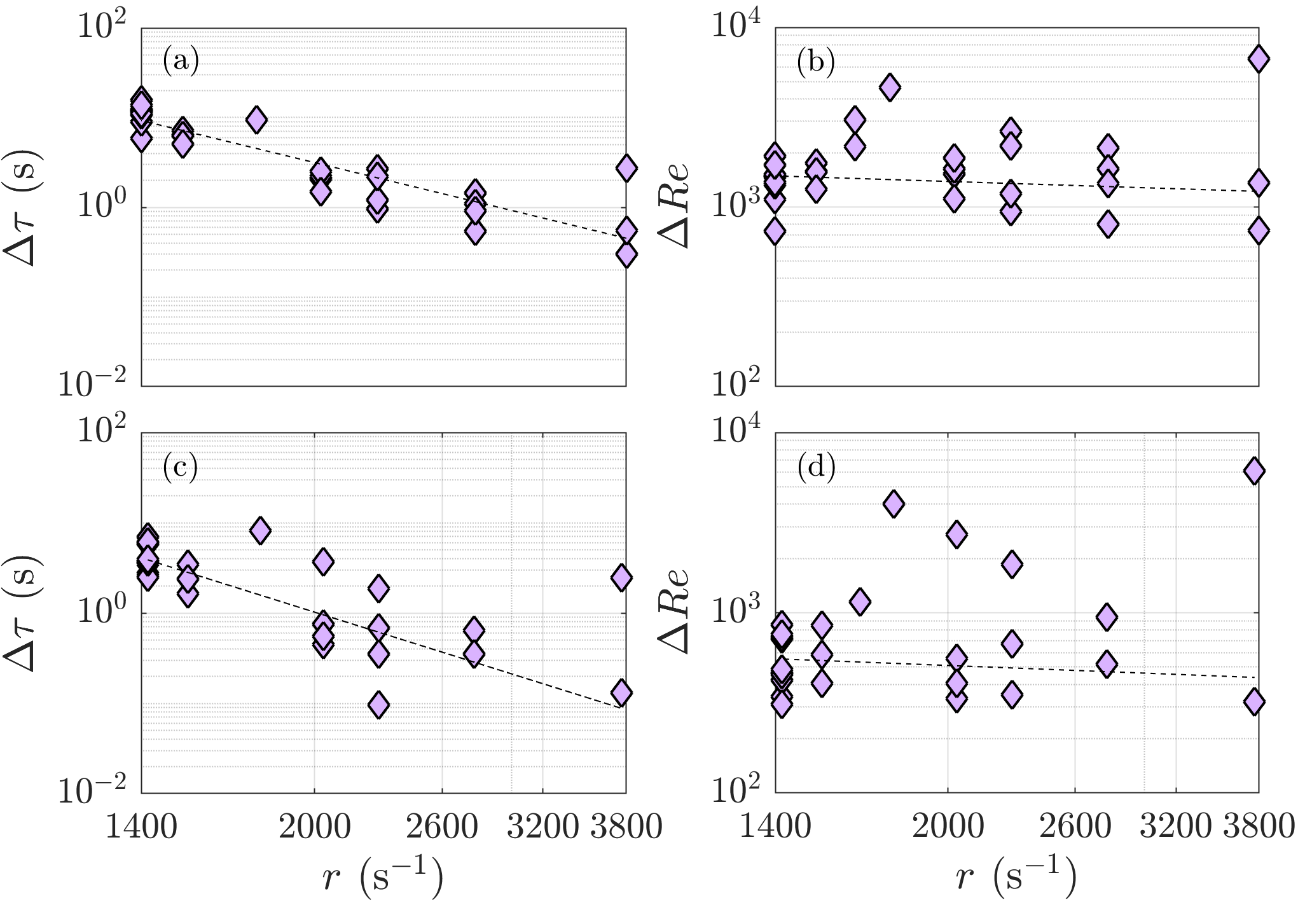}
    \caption{ Warning times ($\Delta \tau$, a-c) and warning parameter intervals $(\Delta Re,b-d)$ are shown by the black diamond marker filled with pink. Here, $\Delta \tau$ and $\Delta Re$ are calculated using threshold $H=0.2$ in (a-b) and $H=0.1$ in (c-d). Both $\Delta t$ and $\Delta Re$ decrease with increasing $r$. 
    Slopes of fitted curves (dashed black lines representing the coefficient $\alpha$) are (a) $-3.1 \pm 0.6$, (b) $-0.2 \pm 0.4$, (c) $-3.9 \pm 1.3$, and (d) $-0.2 \pm 4$.}
    \label{fig:scaling_warning_time}
\end{figure}
After identifying reliable EWS, we look for warning time ($\Delta \tau$) to evade thermoacoustic instability from the combustor. In this regard, we measure $\Delta \tau$ from $H$. $\Delta \tau$ signifies the difference between the onset of periodicity $(t_c ~\text{at}~ Re_c)$ and the time where the Hurst exponent $(H)$ crosses a threshold value. The value of $H$ shows gradual diminution for all the examined values of $r$ and for every realization. The choice of a threshold value is not unique, and scaling can be obtained for different values. In figure~\ref{fig:scaling_warning_time} we plot $\Delta \tau$ together with the warning parameter value $(\Delta Re)$ in $H$, calculated based on threshold values $H=0.1~\text{and}~0.2$, for different ramping rates $(r)$. In figure~\ref{fig:scaling_warning_time}(a) and (b), $\Delta \tau$ and $\Delta Re$ are calculated based on a threshold value of $H=0.2$. Similarly, in figure~\ref{fig:scaling_warning_time}(c) and (d), we calculate $\Delta \tau$ and $\Delta Re$ by setting the threshold value $H=0.1$. We observe that warning times get reduced with increasing $r$ for both of the thresholds values $H=0.1,~\text{and}~0.2$. Interestingly, the average of warning parameter values ($\Delta Re=r \Delta \tau$) for each $r$ also reduces with increasing values of $r$ as evident from figure~\ref{fig:scaling_warning_time}(b) and (d). This is due to the fact that $\Delta \tau \propto r^{\alpha}$, where the exponent $\alpha<-1$ in the observed scenario. Therefore, $\Delta Re$ will change accordingly $r^{\alpha+1}$. The decay in warning in the parameter value $\Delta Re$ is in stark contrast with the earlier results of Pavithran and Sujith~\cite{pavithran2021effect} in a laminar thermoacoustic system, where authors reported $\alpha>-1$. They found that for smaller (larger) values of $r$ or lower (higher) rates, there is more (less) warning time, and the warning interval in the parameter is small (large).

\subsection{Scaling for predicting amplitude of TAI:}
\begin{figure}[t]
    \centering
    \includegraphics[height=!,width=0.45\textwidth]{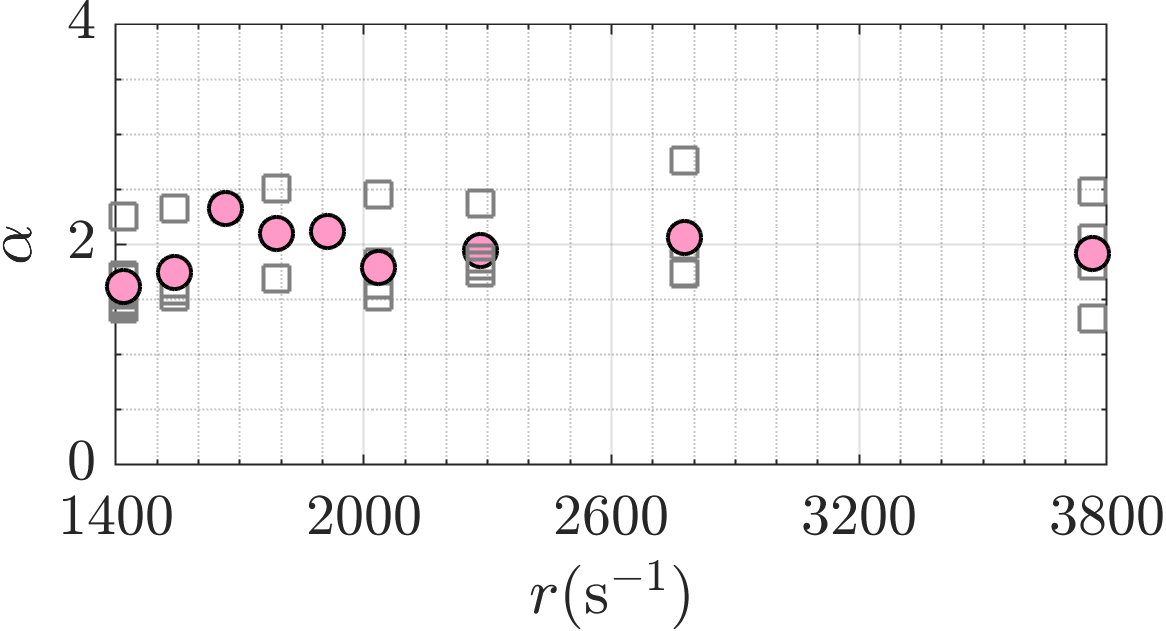}
    \caption{ Exponents of the scaling $A_f \propto H^{-\alpha}$ are plotted (empty squares) in log-log scale for different values of the ramping rate $r$ and for multiple experiments. Empty squares of the same color represent the exponent $\alpha$ for different realizations of an experiment with the same value of $r$. Solid pink circles represent the mean values of $\alpha$. From the figure, it is evident that $\alpha \approx 2\pm0.5$ for $r<3800~\text{s}^{-1}$. We cannot make any conclusive remark about $\alpha$ for higher values of $r$.
   }
    \label{fig:scaling_exponent_alpha}
\end{figure}

Next, we attempt to predict the maximum amplitude - a system can attain during TAI, to understand the severity of TAI and the effect of the rate of change of the control parameter $r$ on the maximum amplitude. To serve the purpose, we plot the absolute value of exponents of the scaling $A_f \propto H^{-\alpha}$, for different values of $r$ in figure~\ref{fig:scaling_exponent_alpha}, where $A_f$ is the most significant mode in the acoustic pressure fluctuations. To compute exponents, we vary $H$ from values nearer to the onset of TAI and continue until the maximum of $A_f$ is attained. Empty squares represent the exponent of different data sets, and colors correspond to different realizations of an experiment. Black circles filled with pink represent mean values. 
For the ramping rate $r<3800~\text{s}^{-1}$ we observe that $\alpha \approx2$, which is consistent with the inverse square law scaling identified by Pavithran \textit{et al.}~\cite{Pavithran_EPL:2020}, in a wide range of systems, with a maximum uncertainty of$0.5$. Due to the lack of data points, it is not possible to make a meaningful calculation of the exponent for ramping rates $r>3000~\text{s}^{-1}$.

\section{Discussions and Conclusions:}

The present work describes our analysis of the robustness and reliability of various early warning signals (EWS) with respect to the rate of change of the control parameter in the context of a turbulent reactive flow system. In this regard, we report the performance of different early warning signals (EWS), based on critical slowing down (CSD), fractal, and spectral characteristics, in detecting thermoacoustic instability (TAI) in a turbulent combustor. The control parameter, the Reynolds number $(Re)$, is changed linearly with time at a finite rate $r$. We also analyze warning times to perform control actions with respect to $r$. Our work reveals that the fractal and spectral-based EWS are more robust and reliable in sensing impending tipping when the underlying complex system has inherent fluctuations and parameters show time dependency. 

In this work, the thermoacoustic system exhibits a transition from a stable chaotic attractor to a stable limit cycle attractor through intermittency as we change $Re$ continuously at a finite rate $r$. Since the state of intermittency comprises random bursts of periodic oscillations among aperiodic epochs, the time series have inherent non-stationarity. On the other hand, statistical indicators such as autocorrelation at lag-$\tau$ and moments of higher orders, i.e., variance, skewness, and kurtosis, are based on the assumption that the underlying time series is statistically stationary. Therefore, in our case, quasi-stationarity is achieved for computing the EWS by selecting an appropriate size of the moving window.

Although characteristics of the CSD phenomenon are smeared by the inherent fluctuations present in the system, an increase in variance and skewness are observed, indicating an impending transition. The increment in skewness is more pronounced for higher values of $r$. Variance, on the other hand, works well for all values of $r$ examined in the present work. This is, perhaps, due to the fact that linear detrending works well to capture fluctuations even in non-stationary time series. 

Earlier works show that the Hurst exponent $(H)$ is a good precursor for non-stationary time series permeated with fluctuations that are deterministic in nature (higher dimensional chaotic nature of turbulent fluctuations in the present context)~\cite{nair_JFM:2014}. Here also, we find that the Hurst exponent gives warnings to an impending TAI. Moreover, warnings due to $H$ are consistent throughout all values of $r$ examined in the present context. Although the lag-1 autocorrelation is permeated with high correlations, the variance of autocorrelation provides a warning of an impending TAI robustly for all examined values of $r$. It will be interesting to investigate how EWS such as the Hurst exponent, variance of autocorrelations, or spectral moments, which are not explicitly quantifying the CSD phenomena, can provide warnings in different ecological or climate systems where the rate of change of a control parameter drives tipping phenomena.

We also explore the reliability of efficient EWS at a wide range of $r$ by revealing their potentiality in providing false alarms. In this context, we compute the area under the curve (AUC) from curves of the receiver operating characteristics (ROC) for all EWS examined in this work. Our analysis reveals that the Hurst exponent, the variance of autocorrelations, the spectral moment of order 2, and the variance of acoustic pressure fluctuations, having higher AUC values, are more reliable EWS in providing a warning. Although skewness possesses a higher AUC value, its ROC curves show that it is not a reliable EWS. Thus, the quantification of an EWS in providing false warnings is very much needed, even though it seems like the indicator is a better EWS.

From the perspective of prevention, we analyze warning times to TAI for different rates. Although warning times decrease with higher values of $r$, we observe that warning in parameter values also decreases in the present context. This observation is in contrast to the laminar case reported earlier \cite{pavithran2021effect}. Finally, the exponents of the scaling of the amplitude $A_f$ with $H$ show consistency with the earlier reported cases of autonomous systems for $r<3800~\text{s}^{-1}$ and take values $2$ with a maximum uncertainty of $\pm0.5$. 
Such kind of scaling is not realized for higher values of $r$ ($r>3800~\text{s}^{-1}$) due to insufficient data points. For high rates, we do not have enough duration of data to estimate the Hurst exponent. In that case, deriving any scaling relation may produce spurious predictions to $A_f$. Therefore, we need more careful observations in order to predict the maxima of $A_f$, or $p_{rms}$ or deriving early warnings for higher values of $r$.

\section*{Acknowledgement:} We express our sincere gratitude to Prof. W. Polifke and T. Komarek of TU Munich, Germany, for sharing the design of the combustor. We acknowledge the help from R. Manikanddan, S. Thilagaraj, S. Anand, and P. R. Midhun during the experiments. The draft has been improved by the fruitful discussions with A. Sahay, and R. Rohit. R.I.S. acknowledges the financial support from the Science and Engineering Research Board (SERB) of the Department of Science and Technology (grant no; CRG/2020/003051).
\appendix
\numberwithin{equation}{section}
\renewcommand{\theequation}{\thesection\arabic{equation}}
\setcounter{equation}{0}

\bibliographystyle{apsrev4-2} 
\bibliography{bibliography_file_EWS}

\begin{thebibliography}{45}%
\makeatletter
\providecommand \@ifxundefined [1]{%
 \@ifx{#1\undefined}
}%
\providecommand \@ifnum [1]{%
 \ifnum #1\expandafter \@firstoftwo
 \else \expandafter \@secondoftwo
 \fi
}%
\providecommand \@ifx [1]{%
 \ifx #1\expandafter \@firstoftwo
 \else \expandafter \@secondoftwo
 \fi
}%
\providecommand \natexlab [1]{#1}%
\providecommand \enquote  [1]{``#1''}%
\providecommand \bibnamefont  [1]{#1}%
\providecommand \bibfnamefont [1]{#1}%
\providecommand \citenamefont [1]{#1}%
\providecommand \href@noop [0]{\@secondoftwo}%
\providecommand \href [0]{\begingroup \@sanitize@url \@href}%
\providecommand \@href[1]{\@@startlink{#1}\@@href}%
\providecommand \@@href[1]{\endgroup#1\@@endlink}%
\providecommand \@sanitize@url [0]{\catcode `\\12\catcode `\$12\catcode
  `\&12\catcode `\#12\catcode `\^12\catcode `\_12\catcode `\%12\relax}%
\providecommand \@@startlink[1]{}%
\providecommand \@@endlink[0]{}%
\providecommand \url  [0]{\begingroup\@sanitize@url \@url }%
\providecommand \@url [1]{\endgroup\@href {#1}{\urlprefix }}%
\providecommand \urlprefix  [0]{URL }%
\providecommand \Eprint [0]{\href }%
\providecommand \doibase [0]{https://doi.org/}%
\providecommand \selectlanguage [0]{\@gobble}%
\providecommand \bibinfo  [0]{\@secondoftwo}%
\providecommand \bibfield  [0]{\@secondoftwo}%
\providecommand \translation [1]{[#1]}%
\providecommand \BibitemOpen [0]{}%
\providecommand \bibitemStop [0]{}%
\providecommand \bibitemNoStop [0]{.\EOS\space}%
\providecommand \EOS [0]{\spacefactor3000\relax}%
\providecommand \BibitemShut  [1]{\csname bibitem#1\endcsname}%
\let\auto@bib@innerbib\@empty
\bibitem [{\citenamefont {van Nes}\ \emph {et~al.}(2016)\citenamefont {van
  Nes}, \citenamefont {Arani}, \citenamefont {Staal}, \citenamefont {van~der
  Bolt}, \citenamefont {Flores}, \citenamefont {Bathiany},\ and\ \citenamefont
  {Scheffer}}]{vanNes:TEE:2016}%
  \BibitemOpen
  \bibfield  {author} {\bibinfo {author} {\bibfnamefont {E.~H.}\ \bibnamefont
  {van Nes}}, \bibinfo {author} {\bibfnamefont {B.~M.}\ \bibnamefont {Arani}},
  \bibinfo {author} {\bibfnamefont {A.}~\bibnamefont {Staal}}, \bibinfo
  {author} {\bibfnamefont {B.}~\bibnamefont {van~der Bolt}}, \bibinfo {author}
  {\bibfnamefont {B.~M.}\ \bibnamefont {Flores}}, \bibinfo {author}
  {\bibfnamefont {S.}~\bibnamefont {Bathiany}},\ and\ \bibinfo {author}
  {\bibfnamefont {M.}~\bibnamefont {Scheffer}},\ }\href@noop {} {\bibfield
  {journal} {\bibinfo  {journal} {Trends in Ecology \& Evolution}\ }\textbf
  {\bibinfo {volume} {31}},\ \bibinfo {pages} {902} (\bibinfo {year}
  {2016})}\BibitemShut {NoStop}%
\bibitem [{\citenamefont {Gladwell}(2006)}]{Gladwell_tipping_book:2006}%
  \BibitemOpen
  \bibfield  {author} {\bibinfo {author} {\bibfnamefont {M.}~\bibnamefont
  {Gladwell}},\ }\href@noop {} {\emph {\bibinfo {title} {The tipping point: How
  little things can make a big difference}}}\ (\bibinfo  {publisher} {Little,
  Brown},\ \bibinfo {year} {2006})\BibitemShut {NoStop}%
\bibitem [{\citenamefont {Dakos}\ \emph {et~al.}(2008)\citenamefont {Dakos},
  \citenamefont {Scheffer}, \citenamefont {van Nes}, \citenamefont {Brovkin},
  \citenamefont {Petoukhov},\ and\ \citenamefont {Held}}]{dakos_PNAS:2008}%
  \BibitemOpen
  \bibfield  {author} {\bibinfo {author} {\bibfnamefont {V.}~\bibnamefont
  {Dakos}}, \bibinfo {author} {\bibfnamefont {M.}~\bibnamefont {Scheffer}},
  \bibinfo {author} {\bibfnamefont {E.~H.}\ \bibnamefont {van Nes}}, \bibinfo
  {author} {\bibfnamefont {V.}~\bibnamefont {Brovkin}}, \bibinfo {author}
  {\bibfnamefont {V.}~\bibnamefont {Petoukhov}},\ and\ \bibinfo {author}
  {\bibfnamefont {H.}~\bibnamefont {Held}},\ }\href@noop {} {\bibfield
  {journal} {\bibinfo  {journal} {Proceedings of the National Academy of
  Sciences}\ }\textbf {\bibinfo {volume} {105}},\ \bibinfo {pages} {14308}
  (\bibinfo {year} {2008})}\BibitemShut {NoStop}%
\bibitem [{\citenamefont {Scheffer}\ \emph {et~al.}(2001)\citenamefont
  {Scheffer}, \citenamefont {Carpenter}, \citenamefont {Foley}, \citenamefont
  {Folke},\ and\ \citenamefont {Walker}}]{scheffer_Nature:2001}%
  \BibitemOpen
  \bibfield  {author} {\bibinfo {author} {\bibfnamefont {M.}~\bibnamefont
  {Scheffer}}, \bibinfo {author} {\bibfnamefont {S.}~\bibnamefont {Carpenter}},
  \bibinfo {author} {\bibfnamefont {J.~A.}\ \bibnamefont {Foley}}, \bibinfo
  {author} {\bibfnamefont {C.}~\bibnamefont {Folke}},\ and\ \bibinfo {author}
  {\bibfnamefont {B.}~\bibnamefont {Walker}},\ }\href@noop {} {\bibfield
  {journal} {\bibinfo  {journal} {Nature}\ }\textbf {\bibinfo {volume} {413}},\
  \bibinfo {pages} {591} (\bibinfo {year} {2001})}\BibitemShut {NoStop}%
\bibitem [{\citenamefont {Pedersen}\ \emph {et~al.}(2017)\citenamefont
  {Pedersen}, \citenamefont {Thompson}, \citenamefont {Ball}, \citenamefont
  {Fortin}, \citenamefont {Gouhier}, \citenamefont {Link}, \citenamefont
  {Moritz}, \citenamefont {Nenzen}, \citenamefont {Stanley}, \citenamefont
  {Taranu} \emph {et~al.}}]{pedersen_RSOS:2017}%
  \BibitemOpen
  \bibfield  {author} {\bibinfo {author} {\bibfnamefont {E.~J.}\ \bibnamefont
  {Pedersen}}, \bibinfo {author} {\bibfnamefont {P.~L.}\ \bibnamefont
  {Thompson}}, \bibinfo {author} {\bibfnamefont {R.~A.}\ \bibnamefont {Ball}},
  \bibinfo {author} {\bibfnamefont {M.-J.}\ \bibnamefont {Fortin}}, \bibinfo
  {author} {\bibfnamefont {T.~C.}\ \bibnamefont {Gouhier}}, \bibinfo {author}
  {\bibfnamefont {H.}~\bibnamefont {Link}}, \bibinfo {author} {\bibfnamefont
  {C.}~\bibnamefont {Moritz}}, \bibinfo {author} {\bibfnamefont
  {H.}~\bibnamefont {Nenzen}}, \bibinfo {author} {\bibfnamefont {R.~R.}\
  \bibnamefont {Stanley}}, \bibinfo {author} {\bibfnamefont {Z.~E.}\
  \bibnamefont {Taranu}}, \emph {et~al.},\ }\href@noop {} {\bibfield  {journal}
  {\bibinfo  {journal} {Royal Society Open Science}\ }\textbf {\bibinfo
  {volume} {4}},\ \bibinfo {pages} {170215} (\bibinfo {year}
  {2017})}\BibitemShut {NoStop}%
\bibitem [{\citenamefont {May}\ \emph {et~al.}(2008)\citenamefont {May},
  \citenamefont {Levin},\ and\ \citenamefont {Sugihara}}]{may_Nature:2008}%
  \BibitemOpen
  \bibfield  {author} {\bibinfo {author} {\bibfnamefont {R.~M.}\ \bibnamefont
  {May}}, \bibinfo {author} {\bibfnamefont {S.~A.}\ \bibnamefont {Levin}},\
  and\ \bibinfo {author} {\bibfnamefont {G.}~\bibnamefont {Sugihara}},\
  }\href@noop {} {\bibfield  {journal} {\bibinfo  {journal} {Nature}\ }\textbf
  {\bibinfo {volume} {451}},\ \bibinfo {pages} {893} (\bibinfo {year}
  {2008})}\BibitemShut {NoStop}%
\bibitem [{\citenamefont {Ren}\ and\ \citenamefont
  {Watts}(2015)}]{ren:EPSR:2015}%
  \BibitemOpen
  \bibfield  {author} {\bibinfo {author} {\bibfnamefont {H.}~\bibnamefont
  {Ren}}\ and\ \bibinfo {author} {\bibfnamefont {D.}~\bibnamefont {Watts}},\
  }\href@noop {} {\bibfield  {journal} {\bibinfo  {journal} {Electric Power
  Systems Research}\ }\textbf {\bibinfo {volume} {124}},\ \bibinfo {pages}
  {173} (\bibinfo {year} {2015})}\BibitemShut {NoStop}%
\bibitem [{\citenamefont {Suchithra}\ \emph {et~al.}(2020)\citenamefont
  {Suchithra}, \citenamefont {Gopalakrishnan}, \citenamefont {Surovyatkina},\
  and\ \citenamefont {Kurths}}]{suchithra:Chaos2020}%
  \BibitemOpen
  \bibfield  {author} {\bibinfo {author} {\bibfnamefont {K.}~\bibnamefont
  {Suchithra}}, \bibinfo {author} {\bibfnamefont {E.}~\bibnamefont
  {Gopalakrishnan}}, \bibinfo {author} {\bibfnamefont {E.}~\bibnamefont
  {Surovyatkina}},\ and\ \bibinfo {author} {\bibfnamefont {J.}~\bibnamefont
  {Kurths}},\ }\href@noop {} {\bibfield  {journal} {\bibinfo  {journal} {Chaos:
  An Interdisciplinary Journal of Nonlinear Science}\ }\textbf {\bibinfo
  {volume} {30}},\ \bibinfo {pages} {061103} (\bibinfo {year}
  {2020})}\BibitemShut {NoStop}%
\bibitem [{\citenamefont {Ivanov}\ \emph {et~al.}(1999)\citenamefont {Ivanov},
  \citenamefont {Amaral}, \citenamefont {Goldberger}, \citenamefont {Havlin},
  \citenamefont {Rosenblum}, \citenamefont {Struzik},\ and\ \citenamefont
  {Stanley}}]{Ivanov_Nature:1999}%
  \BibitemOpen
  \bibfield  {author} {\bibinfo {author} {\bibfnamefont {P.~C.}\ \bibnamefont
  {Ivanov}}, \bibinfo {author} {\bibfnamefont {L.~A.~N.}\ \bibnamefont
  {Amaral}}, \bibinfo {author} {\bibfnamefont {A.~L.}\ \bibnamefont
  {Goldberger}}, \bibinfo {author} {\bibfnamefont {S.}~\bibnamefont {Havlin}},
  \bibinfo {author} {\bibfnamefont {M.~G.}\ \bibnamefont {Rosenblum}}, \bibinfo
  {author} {\bibfnamefont {Z.~R.}\ \bibnamefont {Struzik}},\ and\ \bibinfo
  {author} {\bibfnamefont {H.~E.}\ \bibnamefont {Stanley}},\ }\href@noop {}
  {\bibfield  {journal} {\bibinfo  {journal} {Nature}\ }\textbf {\bibinfo
  {volume} {399}},\ \bibinfo {pages} {461} (\bibinfo {year}
  {1999})}\BibitemShut {NoStop}%
\bibitem [{\citenamefont {McSharry}\ \emph {et~al.}(2003)\citenamefont
  {McSharry}, \citenamefont {Smith},\ and\ \citenamefont
  {Tarassenko}}]{Mcsharry_NatMed:2003}%
  \BibitemOpen
  \bibfield  {author} {\bibinfo {author} {\bibfnamefont {P.~E.}\ \bibnamefont
  {McSharry}}, \bibinfo {author} {\bibfnamefont {L.~A.}\ \bibnamefont
  {Smith}},\ and\ \bibinfo {author} {\bibfnamefont {L.}~\bibnamefont
  {Tarassenko}},\ }\href@noop {} {\bibfield  {journal} {\bibinfo  {journal}
  {Nature Medicine}\ }\textbf {\bibinfo {volume} {9}},\ \bibinfo {pages} {241}
  (\bibinfo {year} {2003})}\BibitemShut {NoStop}%
\bibitem [{\citenamefont {Sarkar}\ \emph {et~al.}(2019)\citenamefont {Sarkar},
  \citenamefont {Sinha}, \citenamefont {Levine}, \citenamefont {Jolly},\ and\
  \citenamefont {Dutta}}]{sarkar_PNAS:2019}%
  \BibitemOpen
  \bibfield  {author} {\bibinfo {author} {\bibfnamefont {S.}~\bibnamefont
  {Sarkar}}, \bibinfo {author} {\bibfnamefont {S.~K.}\ \bibnamefont {Sinha}},
  \bibinfo {author} {\bibfnamefont {H.}~\bibnamefont {Levine}}, \bibinfo
  {author} {\bibfnamefont {M.~K.}\ \bibnamefont {Jolly}},\ and\ \bibinfo
  {author} {\bibfnamefont {P.~S.}\ \bibnamefont {Dutta}},\ }\href@noop {}
  {\bibfield  {journal} {\bibinfo  {journal} {Proceedings of the National
  Academy of Sciences}\ }\textbf {\bibinfo {volume} {116}},\ \bibinfo {pages}
  {26343} (\bibinfo {year} {2019})}\BibitemShut {NoStop}%
\bibitem [{\citenamefont {Ashwin}\ \emph {et~al.}(2012)\citenamefont {Ashwin},
  \citenamefont {Wieczorek}, \citenamefont {Vitolo},\ and\ \citenamefont
  {Cox}}]{ashwin_PhilTracRSA:2012}%
  \BibitemOpen
  \bibfield  {author} {\bibinfo {author} {\bibfnamefont {P.}~\bibnamefont
  {Ashwin}}, \bibinfo {author} {\bibfnamefont {S.}~\bibnamefont {Wieczorek}},
  \bibinfo {author} {\bibfnamefont {R.}~\bibnamefont {Vitolo}},\ and\ \bibinfo
  {author} {\bibfnamefont {P.}~\bibnamefont {Cox}},\ }\href@noop {} {\bibfield
  {journal} {\bibinfo  {journal} {Philosophical Transactions of the Royal
  Society A: Mathematical, Physical and Engineering Sciences}\ }\textbf
  {\bibinfo {volume} {370}},\ \bibinfo {pages} {1166} (\bibinfo {year}
  {2012})}\BibitemShut {NoStop}%
\bibitem [{\citenamefont {Scheffer}\ \emph {et~al.}(2009)\citenamefont
  {Scheffer}, \citenamefont {Bascompte}, \citenamefont {Brock}, \citenamefont
  {Brovkin}, \citenamefont {Carpenter}, \citenamefont {Dakos}, \citenamefont
  {Held}, \citenamefont {Van~Nes}, \citenamefont {Rietkerk},\ and\
  \citenamefont {Sugihara}}]{scheffer_NATURE:2009}%
  \BibitemOpen
  \bibfield  {author} {\bibinfo {author} {\bibfnamefont {M.}~\bibnamefont
  {Scheffer}}, \bibinfo {author} {\bibfnamefont {J.}~\bibnamefont {Bascompte}},
  \bibinfo {author} {\bibfnamefont {W.~A.}\ \bibnamefont {Brock}}, \bibinfo
  {author} {\bibfnamefont {V.}~\bibnamefont {Brovkin}}, \bibinfo {author}
  {\bibfnamefont {S.~R.}\ \bibnamefont {Carpenter}}, \bibinfo {author}
  {\bibfnamefont {V.}~\bibnamefont {Dakos}}, \bibinfo {author} {\bibfnamefont
  {H.}~\bibnamefont {Held}}, \bibinfo {author} {\bibfnamefont {E.~H.}\
  \bibnamefont {Van~Nes}}, \bibinfo {author} {\bibfnamefont {M.}~\bibnamefont
  {Rietkerk}},\ and\ \bibinfo {author} {\bibfnamefont {G.}~\bibnamefont
  {Sugihara}},\ }\href@noop {} {\bibfield  {journal} {\bibinfo  {journal}
  {Nature}\ }\textbf {\bibinfo {volume} {461}},\ \bibinfo {pages} {53}
  (\bibinfo {year} {2009})}\BibitemShut {NoStop}%
\bibitem [{\citenamefont {Strogatz}(2018)}]{strogatz_nonlinear_book:2018}%
  \BibitemOpen
  \bibfield  {author} {\bibinfo {author} {\bibfnamefont {S.~H.}\ \bibnamefont
  {Strogatz}},\ }\href@noop {} {\emph {\bibinfo {title} {Nonlinear Dynamics and
  Chaos with Student Solutions Manual: With Applications to Physics, Biology,
  Chemistry, and Engineering}}}\ (\bibinfo  {publisher} {CRC press},\ \bibinfo
  {year} {2018})\BibitemShut {NoStop}%
\bibitem [{\citenamefont {Guttal}\ and\ \citenamefont
  {Jayaprakash}(2008)}]{guttal_EL:2008}%
  \BibitemOpen
  \bibfield  {author} {\bibinfo {author} {\bibfnamefont {V.}~\bibnamefont
  {Guttal}}\ and\ \bibinfo {author} {\bibfnamefont {C.}~\bibnamefont
  {Jayaprakash}},\ }\href@noop {} {\bibfield  {journal} {\bibinfo  {journal}
  {Ecology Letters}\ }\textbf {\bibinfo {volume} {11}},\ \bibinfo {pages} {450}
  (\bibinfo {year} {2008})}\BibitemShut {NoStop}%
\bibitem [{\citenamefont {Biggs}\ \emph {et~al.}(2009)\citenamefont {Biggs},
  \citenamefont {Carpenter},\ and\ \citenamefont {Brock}}]{Biggs_PNAS:2009}%
  \BibitemOpen
  \bibfield  {author} {\bibinfo {author} {\bibfnamefont {R.}~\bibnamefont
  {Biggs}}, \bibinfo {author} {\bibfnamefont {S.~R.}\ \bibnamefont
  {Carpenter}},\ and\ \bibinfo {author} {\bibfnamefont {W.~A.}\ \bibnamefont
  {Brock}},\ }\href@noop {} {\bibfield  {journal} {\bibinfo  {journal}
  {Proceedings of the National Academy of Sciences}\ }\textbf {\bibinfo
  {volume} {106}},\ \bibinfo {pages} {826} (\bibinfo {year}
  {2009})}\BibitemShut {NoStop}%
\bibitem [{\citenamefont {Bury}\ \emph {et~al.}(2020)\citenamefont {Bury},
  \citenamefont {Bauch},\ and\ \citenamefont {Anand}}]{bury_RSI:2020}%
  \BibitemOpen
  \bibfield  {author} {\bibinfo {author} {\bibfnamefont {T.~M.}\ \bibnamefont
  {Bury}}, \bibinfo {author} {\bibfnamefont {C.~T.}\ \bibnamefont {Bauch}},\
  and\ \bibinfo {author} {\bibfnamefont {M.}~\bibnamefont {Anand}},\
  }\href@noop {} {\bibfield  {journal} {\bibinfo  {journal} {Journal of the
  Royal Society Interface}\ }\textbf {\bibinfo {volume} {17}},\ \bibinfo
  {pages} {20200482} (\bibinfo {year} {2020})}\BibitemShut {NoStop}%
\bibitem [{\citenamefont {Dakos}\ \emph {et~al.}(2012)\citenamefont {Dakos},
  \citenamefont {Carpenter}, \citenamefont {Brock}, \citenamefont {Ellison},
  \citenamefont {Guttal}, \citenamefont {Ives}, \citenamefont {K{\'e}fi},
  \citenamefont {Livina}, \citenamefont {Seekell}, \citenamefont {van Nes}
  \emph {et~al.}}]{dakos_PloSone:2012}%
  \BibitemOpen
  \bibfield  {author} {\bibinfo {author} {\bibfnamefont {V.}~\bibnamefont
  {Dakos}}, \bibinfo {author} {\bibfnamefont {S.~R.}\ \bibnamefont
  {Carpenter}}, \bibinfo {author} {\bibfnamefont {W.~A.}\ \bibnamefont
  {Brock}}, \bibinfo {author} {\bibfnamefont {A.~M.}\ \bibnamefont {Ellison}},
  \bibinfo {author} {\bibfnamefont {V.}~\bibnamefont {Guttal}}, \bibinfo
  {author} {\bibfnamefont {A.~R.}\ \bibnamefont {Ives}}, \bibinfo {author}
  {\bibfnamefont {S.}~\bibnamefont {K{\'e}fi}}, \bibinfo {author}
  {\bibfnamefont {V.}~\bibnamefont {Livina}}, \bibinfo {author} {\bibfnamefont
  {D.~A.}\ \bibnamefont {Seekell}}, \bibinfo {author} {\bibfnamefont {E.~H.}\
  \bibnamefont {van Nes}}, \emph {et~al.},\ }\href@noop {} {\bibfield
  {journal} {\bibinfo  {journal} {PloS one}\ }\textbf {\bibinfo {volume} {7}},\
  \bibinfo {pages} {e41010} (\bibinfo {year} {2012})}\BibitemShut {NoStop}%
\bibitem [{\citenamefont {Pavithran}\ \emph
  {et~al.}(2020{\natexlab{a}})\citenamefont {Pavithran}, \citenamefont {Unni},
  \citenamefont {Varghese}, \citenamefont {Premraj}, \citenamefont {Sujith},
  \citenamefont {Vijayan}, \citenamefont {Saha}, \citenamefont {Marwan},\ and\
  \citenamefont {Kurths}}]{pavithran_SciRep:2020}%
  \BibitemOpen
  \bibfield  {author} {\bibinfo {author} {\bibfnamefont {I.}~\bibnamefont
  {Pavithran}}, \bibinfo {author} {\bibfnamefont {V.~R.}\ \bibnamefont {Unni}},
  \bibinfo {author} {\bibfnamefont {A.~J.}\ \bibnamefont {Varghese}}, \bibinfo
  {author} {\bibfnamefont {D.}~\bibnamefont {Premraj}}, \bibinfo {author}
  {\bibfnamefont {R.~I.}\ \bibnamefont {Sujith}}, \bibinfo {author}
  {\bibfnamefont {C.}~\bibnamefont {Vijayan}}, \bibinfo {author} {\bibfnamefont
  {A.}~\bibnamefont {Saha}}, \bibinfo {author} {\bibfnamefont {N.}~\bibnamefont
  {Marwan}},\ and\ \bibinfo {author} {\bibfnamefont {J.}~\bibnamefont
  {Kurths}},\ }\href@noop {} {\bibfield  {journal} {\bibinfo  {journal}
  {Scientific Reports}\ }\textbf {\bibinfo {volume} {10}},\ \bibinfo {pages}
  {1} (\bibinfo {year} {2020}{\natexlab{a}})}\BibitemShut {NoStop}%
\bibitem [{\citenamefont {Lieuwen}(2021)}]{lieuwen2021unsteady}%
  \BibitemOpen
  \bibfield  {author} {\bibinfo {author} {\bibfnamefont {T.~C.}\ \bibnamefont
  {Lieuwen}},\ }\href@noop {} {\emph {\bibinfo {title} {Unsteady Combustor
  Physics}}}\ (\bibinfo  {publisher} {Cambridge University Press},\ \bibinfo
  {year} {2021})\BibitemShut {NoStop}%
\bibitem [{\citenamefont {Sujith}\ and\ \citenamefont
  {Pawar}(2021)}]{sujith_TI:2021}%
  \BibitemOpen
  \bibfield  {author} {\bibinfo {author} {\bibfnamefont {R.~I.}\ \bibnamefont
  {Sujith}}\ and\ \bibinfo {author} {\bibfnamefont {S.~A.}\ \bibnamefont
  {Pawar}},\ }\href@noop {} {\bibfield  {journal} {\bibinfo  {journal}
  {Springer Series in Synergetics}\ } (\bibinfo {year} {2021})}\BibitemShut
  {NoStop}%
\bibitem [{\citenamefont {Jegadeesan}\ and\ \citenamefont
  {Sujith}(2013)}]{jegadeesan_ntipping:2013}%
  \BibitemOpen
  \bibfield  {author} {\bibinfo {author} {\bibfnamefont {V.}~\bibnamefont
  {Jegadeesan}}\ and\ \bibinfo {author} {\bibfnamefont {R.~I.}\ \bibnamefont
  {Sujith}},\ }\href@noop {} {\bibfield  {journal} {\bibinfo  {journal}
  {Proceedings of the Combustion Institute}\ }\textbf {\bibinfo {volume}
  {34}},\ \bibinfo {pages} {3175} (\bibinfo {year} {2013})}\BibitemShut
  {NoStop}%
\bibitem [{\citenamefont {Manikandan}\ and\ \citenamefont
  {Sujith}(2020)}]{manikandan_afterburner:2020}%
  \BibitemOpen
  \bibfield  {author} {\bibinfo {author} {\bibfnamefont {S.}~\bibnamefont
  {Manikandan}}\ and\ \bibinfo {author} {\bibfnamefont {R.~I.}\ \bibnamefont
  {Sujith}},\ }\href@noop {} {\bibfield  {journal} {\bibinfo  {journal}
  {Experimental Thermal and Fluid Science}\ }\textbf {\bibinfo {volume}
  {114}},\ \bibinfo {pages} {110046} (\bibinfo {year} {2020})}\BibitemShut
  {NoStop}%
\bibitem [{\citenamefont {Nair}\ \emph {et~al.}(2013)\citenamefont {Nair},
  \citenamefont {Thampi}, \citenamefont {Karuppusamy}, \citenamefont
  {Gopalan},\ and\ \citenamefont {Sujith}}]{nair_IJSCD:2013}%
  \BibitemOpen
  \bibfield  {author} {\bibinfo {author} {\bibfnamefont {V.}~\bibnamefont
  {Nair}}, \bibinfo {author} {\bibfnamefont {G.}~\bibnamefont {Thampi}},
  \bibinfo {author} {\bibfnamefont {S.}~\bibnamefont {Karuppusamy}}, \bibinfo
  {author} {\bibfnamefont {S.}~\bibnamefont {Gopalan}},\ and\ \bibinfo {author}
  {\bibfnamefont {R.~I.}\ \bibnamefont {Sujith}},\ }\href@noop {} {\bibfield
  {journal} {\bibinfo  {journal} {International Journal of Spray and Combustion
  Dynamics}\ }\textbf {\bibinfo {volume} {5}},\ \bibinfo {pages} {273}
  (\bibinfo {year} {2013})}\BibitemShut {NoStop}%
\bibitem [{\citenamefont {Tony}\ \emph {et~al.}(2015)\citenamefont {Tony},
  \citenamefont {Gopalakrishnan}, \citenamefont {Sreelekha},\ and\
  \citenamefont {Sujith}}]{tony_PRE:2015}%
  \BibitemOpen
  \bibfield  {author} {\bibinfo {author} {\bibfnamefont {J.}~\bibnamefont
  {Tony}}, \bibinfo {author} {\bibfnamefont {E.~A.}\ \bibnamefont
  {Gopalakrishnan}}, \bibinfo {author} {\bibfnamefont {E.}~\bibnamefont
  {Sreelekha}},\ and\ \bibinfo {author} {\bibfnamefont {R.~I.}\ \bibnamefont
  {Sujith}},\ }\href@noop {} {\bibfield  {journal} {\bibinfo  {journal} {Phys.
  Rev. E}\ }\textbf {\bibinfo {volume} {92}},\ \bibinfo {pages} {062902}
  (\bibinfo {year} {2015})}\BibitemShut {NoStop}%
\bibitem [{\citenamefont {Tony}\ \emph {et~al.}(2017)\citenamefont {Tony},
  \citenamefont {Subarna}, \citenamefont {Syamkumar}, \citenamefont {Sudha},
  \citenamefont {Akshay}, \citenamefont {Gopalakrishnan}, \citenamefont
  {Surovyatkina},\ and\ \citenamefont {Sujith}}]{tony_SciRep:2017}%
  \BibitemOpen
  \bibfield  {author} {\bibinfo {author} {\bibfnamefont {J.}~\bibnamefont
  {Tony}}, \bibinfo {author} {\bibfnamefont {S.}~\bibnamefont {Subarna}},
  \bibinfo {author} {\bibfnamefont {K.~S.}\ \bibnamefont {Syamkumar}}, \bibinfo
  {author} {\bibfnamefont {G.}~\bibnamefont {Sudha}}, \bibinfo {author}
  {\bibfnamefont {S.}~\bibnamefont {Akshay}}, \bibinfo {author} {\bibfnamefont
  {E.~A.}\ \bibnamefont {Gopalakrishnan}}, \bibinfo {author} {\bibfnamefont
  {E.}~\bibnamefont {Surovyatkina}},\ and\ \bibinfo {author} {\bibfnamefont
  {R.~I.}\ \bibnamefont {Sujith}},\ }\href@noop {} {\bibfield  {journal}
  {\bibinfo  {journal} {Scientific Reports}\ }\textbf {\bibinfo {volume} {7}},\
  \bibinfo {pages} {1} (\bibinfo {year} {2017})}\BibitemShut {NoStop}%
\bibitem [{\citenamefont {Bonciolini}\ \emph {et~al.}(2018)\citenamefont
  {Bonciolini}, \citenamefont {Ebi}, \citenamefont {Boujo},\ and\ \citenamefont
  {Noiray}}]{bonciolini_RSOS:2018}%
  \BibitemOpen
  \bibfield  {author} {\bibinfo {author} {\bibfnamefont {G.}~\bibnamefont
  {Bonciolini}}, \bibinfo {author} {\bibfnamefont {D.}~\bibnamefont {Ebi}},
  \bibinfo {author} {\bibfnamefont {E.}~\bibnamefont {Boujo}},\ and\ \bibinfo
  {author} {\bibfnamefont {N.}~\bibnamefont {Noiray}},\ }\href@noop {}
  {\bibfield  {journal} {\bibinfo  {journal} {Royal Society Open Science}\
  }\textbf {\bibinfo {volume} {5}},\ \bibinfo {pages} {172078} (\bibinfo {year}
  {2018})}\BibitemShut {NoStop}%
\bibitem [{\citenamefont {Unni}\ \emph {et~al.}(2019)\citenamefont {Unni},
  \citenamefont {Gopalakrishnan}, \citenamefont {Syamkumar}, \citenamefont
  {Sujith}, \citenamefont {Surovyatkina},\ and\ \citenamefont
  {Kurths}}]{unni_Chaos:2019}%
  \BibitemOpen
  \bibfield  {author} {\bibinfo {author} {\bibfnamefont {V.~R.}\ \bibnamefont
  {Unni}}, \bibinfo {author} {\bibfnamefont {E.~A.}\ \bibnamefont
  {Gopalakrishnan}}, \bibinfo {author} {\bibfnamefont {K.~S.}\ \bibnamefont
  {Syamkumar}}, \bibinfo {author} {\bibfnamefont {R.~I.}\ \bibnamefont
  {Sujith}}, \bibinfo {author} {\bibfnamefont {E.}~\bibnamefont
  {Surovyatkina}},\ and\ \bibinfo {author} {\bibfnamefont {J.}~\bibnamefont
  {Kurths}},\ }\href@noop {} {\bibfield  {journal} {\bibinfo  {journal} {Chaos:
  An Interdisciplinary Journal of Nonlinear Science}\ }\textbf {\bibinfo
  {volume} {29}},\ \bibinfo {pages} {031102} (\bibinfo {year}
  {2019})}\BibitemShut {NoStop}%
\bibitem [{\citenamefont {Gopalakrishnan}\ \emph {et~al.}(2016)\citenamefont
  {Gopalakrishnan}, \citenamefont {Sharma}, \citenamefont {John}, \citenamefont
  {Dutta},\ and\ \citenamefont {Sujith}}]{gopalakrishnan_SciRep:2016}%
  \BibitemOpen
  \bibfield  {author} {\bibinfo {author} {\bibfnamefont {E.~A.}\ \bibnamefont
  {Gopalakrishnan}}, \bibinfo {author} {\bibfnamefont {Y.}~\bibnamefont
  {Sharma}}, \bibinfo {author} {\bibfnamefont {T.}~\bibnamefont {John}},
  \bibinfo {author} {\bibfnamefont {P.~S.}\ \bibnamefont {Dutta}},\ and\
  \bibinfo {author} {\bibfnamefont {R.~I.}\ \bibnamefont {Sujith}},\
  }\href@noop {} {\bibfield  {journal} {\bibinfo  {journal} {Scientific
  Reports}\ }\textbf {\bibinfo {volume} {6}},\ \bibinfo {pages} {1} (\bibinfo
  {year} {2016})}\BibitemShut {NoStop}%
\bibitem [{\citenamefont {An}\ \emph {et~al.}(2019)\citenamefont {An},
  \citenamefont {Steinberg}, \citenamefont {Jella}, \citenamefont {Bourque},\
  and\ \citenamefont {F{\"u}ri}}]{An_JEGTP:2019}%
  \BibitemOpen
  \bibfield  {author} {\bibinfo {author} {\bibfnamefont {Q.}~\bibnamefont
  {An}}, \bibinfo {author} {\bibfnamefont {A.~M.}\ \bibnamefont {Steinberg}},
  \bibinfo {author} {\bibfnamefont {S.}~\bibnamefont {Jella}}, \bibinfo
  {author} {\bibfnamefont {G.}~\bibnamefont {Bourque}},\ and\ \bibinfo {author}
  {\bibfnamefont {M.}~\bibnamefont {F{\"u}ri}},\ }\href@noop {} {\bibfield
  {journal} {\bibinfo  {journal} {Journal of Engineering for Gas Turbines and
  Power}\ }\textbf {\bibinfo {volume} {141}} (\bibinfo {year}
  {2019})}\BibitemShut {NoStop}%
\bibitem [{\citenamefont {Pavithran}\ \emph {et~al.}(2021)\citenamefont
  {Pavithran}, \citenamefont {Unni},\ and\ \citenamefont
  {Sujith}}]{pavithran_review_EPJST:2021}%
  \BibitemOpen
  \bibfield  {author} {\bibinfo {author} {\bibfnamefont {I.}~\bibnamefont
  {Pavithran}}, \bibinfo {author} {\bibfnamefont {V.~R.}\ \bibnamefont
  {Unni}},\ and\ \bibinfo {author} {\bibfnamefont {R.~I.}\ \bibnamefont
  {Sujith}},\ }\href@noop {} {\bibfield  {journal} {\bibinfo  {journal} {The
  European Physical Journal Special Topics}\ }\textbf {\bibinfo {volume}
  {230}},\ \bibinfo {pages} {3411} (\bibinfo {year} {2021})}\BibitemShut
  {NoStop}%
\bibitem [{\citenamefont {Pavithran}\ and\ \citenamefont
  {Sujith}(2021)}]{pavithran2021effect}%
  \BibitemOpen
  \bibfield  {author} {\bibinfo {author} {\bibfnamefont {I.}~\bibnamefont
  {Pavithran}}\ and\ \bibinfo {author} {\bibfnamefont {R.~I.}\ \bibnamefont
  {Sujith}},\ }\href@noop {} {\bibfield  {journal} {\bibinfo  {journal} {Chaos:
  An Interdisciplinary Journal of Nonlinear Science}\ }\textbf {\bibinfo
  {volume} {31}},\ \bibinfo {pages} {013116} (\bibinfo {year}
  {2021})}\BibitemShut {NoStop}%
\bibitem [{\citenamefont {Komarek}\ and\ \citenamefont
  {Polifke}(2010)}]{komarek_combustor_design:2010}%
  \BibitemOpen
  \bibfield  {author} {\bibinfo {author} {\bibfnamefont {T.}~\bibnamefont
  {Komarek}}\ and\ \bibinfo {author} {\bibfnamefont {W.}~\bibnamefont
  {Polifke}},\ }\href@noop {} {\bibfield  {journal} {\bibinfo  {journal}
  {Journal of Engineering for Gas Turbines and Power}\ }\textbf {\bibinfo
  {volume} {132}} (\bibinfo {year} {2010})}\BibitemShut {NoStop}%
\bibitem [{\citenamefont {Raghunathan}\ \emph {et~al.}(2020)\citenamefont
  {Raghunathan}, \citenamefont {George}, \citenamefont {Unni}, \citenamefont
  {Midhun}, \citenamefont {Reeja},\ and\ \citenamefont
  {Sujith}}]{raghunathan_JFM:2020}%
  \BibitemOpen
  \bibfield  {author} {\bibinfo {author} {\bibfnamefont {M.}~\bibnamefont
  {Raghunathan}}, \bibinfo {author} {\bibfnamefont {N.~B.}\ \bibnamefont
  {George}}, \bibinfo {author} {\bibfnamefont {V.~R.}\ \bibnamefont {Unni}},
  \bibinfo {author} {\bibfnamefont {P.~R.}\ \bibnamefont {Midhun}}, \bibinfo
  {author} {\bibfnamefont {K.~V.}\ \bibnamefont {Reeja}},\ and\ \bibinfo
  {author} {\bibfnamefont {R.~I.}\ \bibnamefont {Sujith}},\ }\href@noop {}
  {\bibfield  {journal} {\bibinfo  {journal} {Journal of Fluid Mechanics}\
  }\textbf {\bibinfo {volume} {888}} (\bibinfo {year} {2020})}\BibitemShut
  {NoStop}%
\bibitem [{\citenamefont {Nair}\ \emph {et~al.}(2014)\citenamefont {Nair},
  \citenamefont {Thampi},\ and\ \citenamefont {Sujith}}]{nair_thampi_JFM:2014}%
  \BibitemOpen
  \bibfield  {author} {\bibinfo {author} {\bibfnamefont {V.}~\bibnamefont
  {Nair}}, \bibinfo {author} {\bibfnamefont {G.}~\bibnamefont {Thampi}},\ and\
  \bibinfo {author} {\bibfnamefont {R.~I.}\ \bibnamefont {Sujith}},\
  }\href@noop {} {\bibfield  {journal} {\bibinfo  {journal} {Journal of Fluid
  Mechanics}\ }\textbf {\bibinfo {volume} {756}},\ \bibinfo {pages} {470}
  (\bibinfo {year} {2014})}\BibitemShut {NoStop}%
\bibitem [{\citenamefont {Subramanian}\ \emph {et~al.}(2010)\citenamefont
  {Subramanian}, \citenamefont {Mariappan}, \citenamefont {Sujith},\ and\
  \citenamefont {Wahi}}]{subramanian_IJSCD:2010}%
  \BibitemOpen
  \bibfield  {author} {\bibinfo {author} {\bibfnamefont {P.}~\bibnamefont
  {Subramanian}}, \bibinfo {author} {\bibfnamefont {S.}~\bibnamefont
  {Mariappan}}, \bibinfo {author} {\bibfnamefont {R.~I.}\ \bibnamefont
  {Sujith}},\ and\ \bibinfo {author} {\bibfnamefont {P.}~\bibnamefont {Wahi}},\
  }\href@noop {} {\bibfield  {journal} {\bibinfo  {journal} {International
  Journal of Spray and Combustion Dynamics}\ }\textbf {\bibinfo {volume} {2}},\
  \bibinfo {pages} {325} (\bibinfo {year} {2010})}\BibitemShut {NoStop}%
\bibitem [{\citenamefont {Etikyala}\ and\ \citenamefont
  {Sujith}(2017)}]{etikyala_Chaos:2017}%
  \BibitemOpen
  \bibfield  {author} {\bibinfo {author} {\bibfnamefont {S.}~\bibnamefont
  {Etikyala}}\ and\ \bibinfo {author} {\bibfnamefont {R.~I.}\ \bibnamefont
  {Sujith}},\ }\href@noop {} {\bibfield  {journal} {\bibinfo  {journal} {Chaos:
  An Interdisciplinary Journal of Nonlinear Science}\ }\textbf {\bibinfo
  {volume} {27}},\ \bibinfo {pages} {023106} (\bibinfo {year}
  {2017})}\BibitemShut {NoStop}%
\bibitem [{\citenamefont {Bonciolini}\ and\ \citenamefont
  {Noiray}(2019)}]{bonciolini_NonDyn:2019}%
  \BibitemOpen
  \bibfield  {author} {\bibinfo {author} {\bibfnamefont {G.}~\bibnamefont
  {Bonciolini}}\ and\ \bibinfo {author} {\bibfnamefont {N.}~\bibnamefont
  {Noiray}},\ }\href@noop {} {\bibfield  {journal} {\bibinfo  {journal}
  {Nonlinear Dynamics}\ }\textbf {\bibinfo {volume} {96}},\ \bibinfo {pages}
  {703} (\bibinfo {year} {2019})}\BibitemShut {NoStop}%
\bibitem [{\citenamefont {Nair}\ and\ \citenamefont
  {Sujith}(2014)}]{nair_JFM:2014}%
  \BibitemOpen
  \bibfield  {author} {\bibinfo {author} {\bibfnamefont {V.}~\bibnamefont
  {Nair}}\ and\ \bibinfo {author} {\bibfnamefont {R.~I.}\ \bibnamefont
  {Sujith}},\ }\href@noop {} {\bibfield  {journal} {\bibinfo  {journal}
  {Journal of Fluid Mechanics}\ }\textbf {\bibinfo {volume} {747}},\ \bibinfo
  {pages} {635} (\bibinfo {year} {2014})}\BibitemShut {NoStop}%
\bibitem [{\citenamefont {Lilly}\ and\ \citenamefont
  {Olhede}(2008)}]{lilly_Morse_wavelets:2008}%
  \BibitemOpen
  \bibfield  {author} {\bibinfo {author} {\bibfnamefont {J.~M.}\ \bibnamefont
  {Lilly}}\ and\ \bibinfo {author} {\bibfnamefont {S.~C.}\ \bibnamefont
  {Olhede}},\ }\href@noop {} {\bibfield  {journal} {\bibinfo  {journal} {IEEE
  Transactions on Signal Processing}\ }\textbf {\bibinfo {volume} {57}},\
  \bibinfo {pages} {146} (\bibinfo {year} {2008})}\BibitemShut {NoStop}%
\bibitem [{\citenamefont {Calvet}\ and\ \citenamefont
  {Fisher}(2002)}]{calvet_multifractality:2002}%
  \BibitemOpen
  \bibfield  {author} {\bibinfo {author} {\bibfnamefont {L.}~\bibnamefont
  {Calvet}}\ and\ \bibinfo {author} {\bibfnamefont {A.}~\bibnamefont
  {Fisher}},\ }\href@noop {} {\bibfield  {journal} {\bibinfo  {journal} {Review
  of Economics and Statistics}\ }\textbf {\bibinfo {volume} {84}},\ \bibinfo
  {pages} {381} (\bibinfo {year} {2002})}\BibitemShut {NoStop}%
\bibitem [{\citenamefont {Keller}\ \emph {et~al.}(2009)\citenamefont {Keller},
  \citenamefont {Lodge}, \citenamefont {Lewis},\ and\ \citenamefont
  {Shogren}}]{keller2009bioeconomics}%
  \BibitemOpen
  \bibfield  {author} {\bibinfo {author} {\bibfnamefont {R.~P.}\ \bibnamefont
  {Keller}}, \bibinfo {author} {\bibfnamefont {D.~M.}\ \bibnamefont {Lodge}},
  \bibinfo {author} {\bibfnamefont {M.~A.}\ \bibnamefont {Lewis}},\ and\
  \bibinfo {author} {\bibfnamefont {J.~F.}\ \bibnamefont {Shogren}},\
  }\href@noop {} {\emph {\bibinfo {title} {Bioeconomics of Invasive Species:
  Integrating Ecology, Economics, Policy, and Management}}}\ (\bibinfo
  {publisher} {Oxford University Press},\ \bibinfo {year} {2009})\BibitemShut
  {NoStop}%
\bibitem [{\citenamefont {Green}\ \emph {et~al.}(1966)\citenamefont {Green},
  \citenamefont {Swets} \emph {et~al.}}]{green1966signal}%
  \BibitemOpen
  \bibfield  {author} {\bibinfo {author} {\bibfnamefont {D.~M.}\ \bibnamefont
  {Green}}, \bibinfo {author} {\bibfnamefont {J.~A.}\ \bibnamefont {Swets}},
  \emph {et~al.},\ }\href@noop {} {\emph {\bibinfo {title} {Signal Detection
  Theory and Psychophysics}}},\ Vol.~\bibinfo {volume} {1}\ (\bibinfo
  {publisher} {Wiley New York},\ \bibinfo {year} {1966})\BibitemShut {NoStop}%
\bibitem [{\citenamefont {Boettiger}\ and\ \citenamefont
  {Hastings}(2012)}]{boettiger_RSIf:2012}%
  \BibitemOpen
  \bibfield  {author} {\bibinfo {author} {\bibfnamefont {C.}~\bibnamefont
  {Boettiger}}\ and\ \bibinfo {author} {\bibfnamefont {A.}~\bibnamefont
  {Hastings}},\ }\href@noop {} {\bibfield  {journal} {\bibinfo  {journal}
  {Journal of the Royal Society Interface}\ }\textbf {\bibinfo {volume} {9}},\
  \bibinfo {pages} {2527} (\bibinfo {year} {2012})}\BibitemShut {NoStop}%
\bibitem [{\citenamefont {Pavithran}\ \emph
  {et~al.}(2020{\natexlab{b}})\citenamefont {Pavithran}, \citenamefont {Unni},
  \citenamefont {Varghese}, \citenamefont {Sujith}, \citenamefont {Saha},
  \citenamefont {Marwan},\ and\ \citenamefont {Kurths}}]{Pavithran_EPL:2020}%
  \BibitemOpen
  \bibfield  {author} {\bibinfo {author} {\bibfnamefont {I.}~\bibnamefont
  {Pavithran}}, \bibinfo {author} {\bibfnamefont {V.~R.}\ \bibnamefont {Unni}},
  \bibinfo {author} {\bibfnamefont {A.~J.}\ \bibnamefont {Varghese}}, \bibinfo
  {author} {\bibfnamefont {R.~I.}\ \bibnamefont {Sujith}}, \bibinfo {author}
  {\bibfnamefont {A.}~\bibnamefont {Saha}}, \bibinfo {author} {\bibfnamefont
  {N.}~\bibnamefont {Marwan}},\ and\ \bibinfo {author} {\bibfnamefont
  {J.}~\bibnamefont {Kurths}},\ }\href@noop {} {\bibfield  {journal} {\bibinfo
  {journal} {EPL (Europhysics Letters)}\ }\textbf {\bibinfo {volume} {129}},\
  \bibinfo {pages} {24004} (\bibinfo {year} {2020}{\natexlab{b}})}\BibitemShut
  {NoStop}%
\end{thebibliography}%

\end{document}



\title{
Supplementary Material: Early warnings of tipping in a non-autonomous turbulent reactive flow system: efficacy, reliability, and warning times}
\author{Ankan Banerjee}\email{ankan1090@gmail.com}
\affiliation{Department of Aerospace Engineering, Indian Institute of Technology Madras, Chennai 600036, India}  
\author{Induja Pavithran}%
\email{indujap2013@gmail.com}
\affiliation{Department of Aerospace Engineering, Indian Institute of Technology Madras, Chennai 600036, India}
\author{R. I. Sujith}%
\email{sujith@iitm.ac.in}
\affiliation{Department of Aerospace Engineering, Indian Institute of Technology Madras, Chennai 600036, India}%

\date{\today}

\maketitle

\section{Determining size of the moving window:}
\begin{figure*}[b]
    \centering
    \includegraphics[height=!,width=0.88\textwidth]{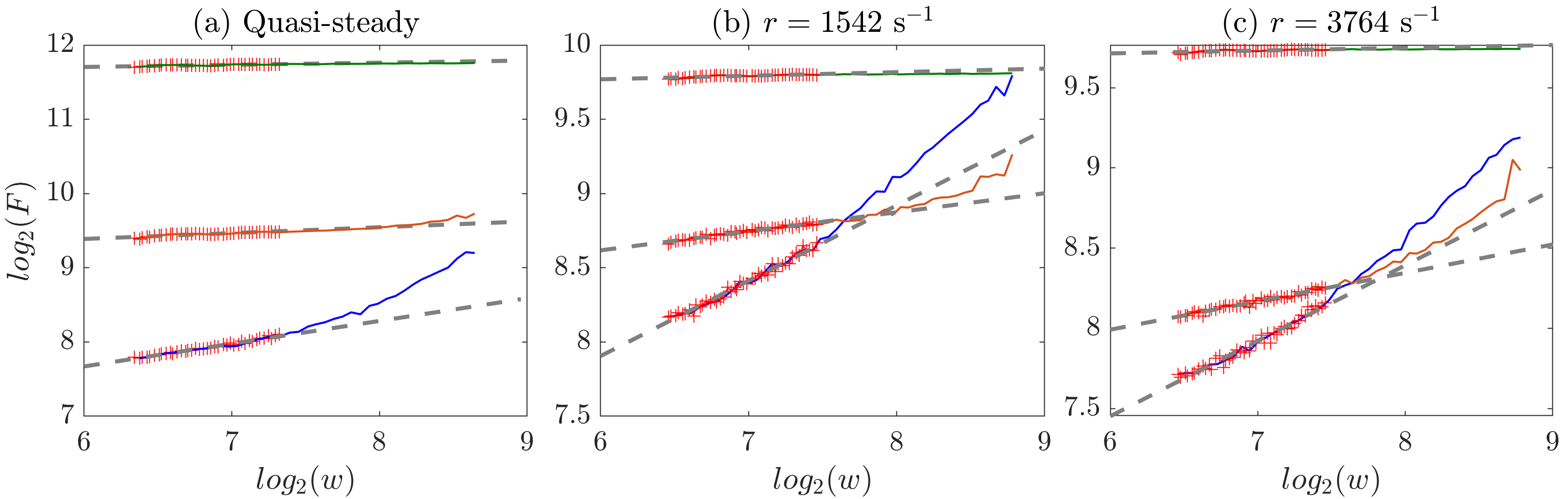}
    \caption{Structure-function ($F$) is plotted for different values of scale ($w$) for (a) autonomous and (b-c) non-autonomous thermoacoustic systems. Rates of change of the control parameter are (b) $r=1542$ s$^{-1}$, and (c) $r=3764$ s$^{-1}$. Blue, orange, and green curves represent the states of combustion noise, intermittency, and thermoacoustic instability, respectively. Red markers represent values of $F$ for scales corresponding to 2-4 acoustic cycles. Gray dashed lines are the slope of the curves whose gradient is the Hurst exponent.}.
    \label{fig:convergence_exponent}
\end{figure*}

We determine the length of a moving window in such a manner that different early warning signals (EWS) at each dynamical state of the underlying system converge to a fixed value. Since the computation of the Hurst exponent ($H$) requires the calculation of small-scale and large-scale fluctuations, so we analyze the convergence of $H$ to identify an appropriate window length. In this regard, we first estimate the maximum and minimum possible length scales ($w$). In figure \ref{fig:convergence_exponent} we plot the structure function of order $2$ ($F$) with respect to the binarized length scale $w$, where
\begin{equation}
    F=\sqrt{1/n\sum_{i=1}^{n} \left(1/w\sum_{t=1}^{w}(p_i(t)-\Bar{p_i})^2\right)},
\end{equation}
where $n$ is the non-overlapping segments of the time series $p(t)$ of equal length $w$, and $\Bar{p_i}$ is the local mean at each of these segments. The linear trend of this curve is the Hurst exponent~\cite{nair_JFM:2014}. The fundamental frequency of the combustor is approximately $275$ Hz, and the sampling frequency is $12$ kHz. We vary scales from 2 to 10 acoustic cycles to compute $H$. Figure \ref{fig:convergence_exponent} shows that the slope of the plot is obtained for 2-4 acoustic cycles at different states corresponding to combustion noise, intermittency, and thermoacoustic instability for both autonomous and non-autonomous thermoacoustic systems.

 \begin{figure*}[h]
    \centering
    \includegraphics[height=!,width=0.88\textwidth]{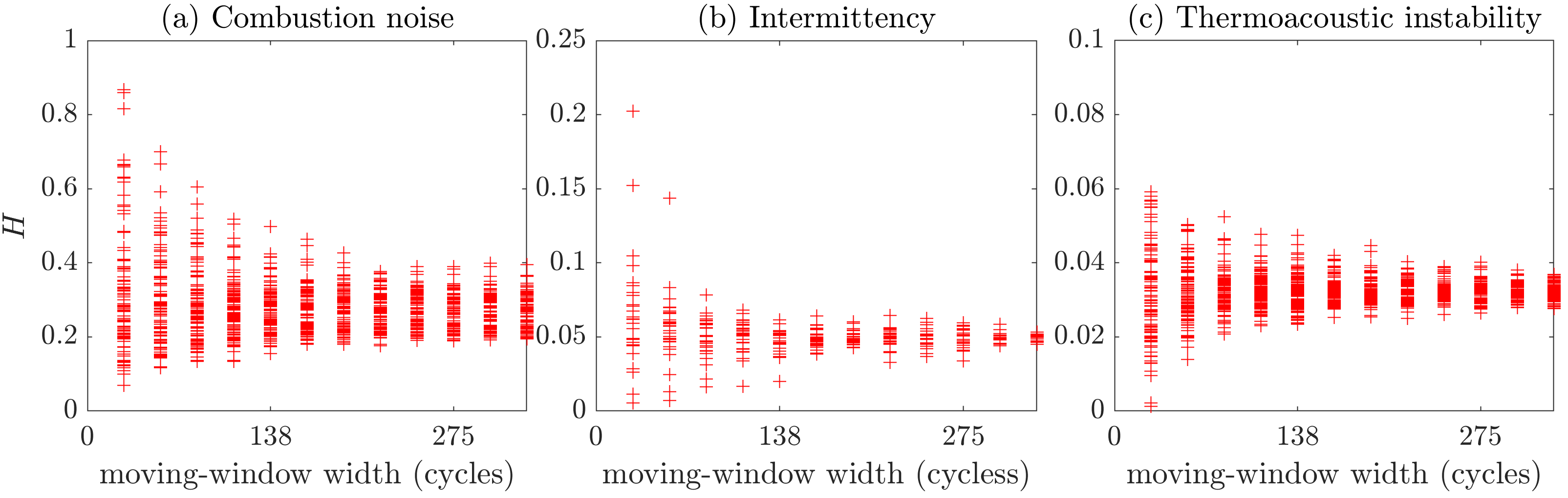}
    \caption{ The Hurst exponent corresponding to a time series of acoustic pressure fluctuations of an autonomous turbulent thermoacoustic system is shown here. Values of $H$ are computed for different sizes of moving windows for the states of (a) combustion noise, (b) intermittency, and (c) thermoacoustic instability. The Hurst exponent tends to a fixed value with a smaller variance for moving-$w$ width higher than 0.6 seconds.}
    \label{fig:convergence_window_size}
\end{figure*}
Once the length scales are fixed then, we compute $H$ for different sizes of a moving window. We measure the width of a moving window in terms of acoustic cycles. Since the fundamental frequency of the combustor we examined is approximately $275$ Hz, a one-second window contains 275 cycles. It is evident from the figure \ref{fig:convergence_window_size} that values of $H$ corresponding to other states of the system converge for window sizes containing 165 or more cycles. We select the minimum possible number of acoustic cycles, which is 165, for calculating the length of the moving window.  Hence, 0.6 seconds is the length of a moving window. Note that we test the convergence of $H$ for quasi-steady time series obtained from the same combustor.

\section{Convergence of the variance of autocorrelation ($var(AC)$):}
\begin{figure}[ht]
    \centering
    \includegraphics[height=!,width=0.5\textwidth]{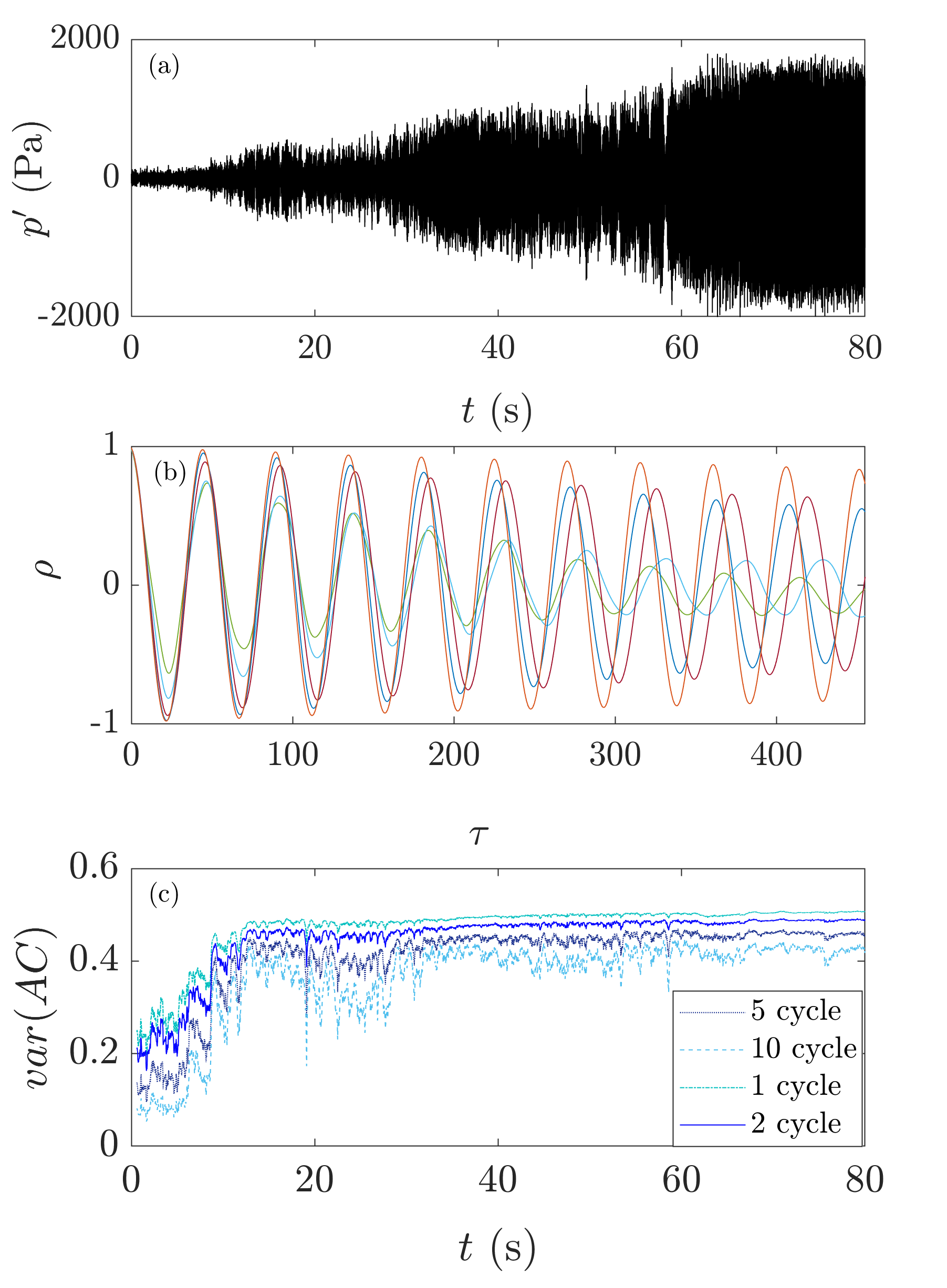}
    \caption{ (a) Time series of acoustic pressure fluctuations ($p^\prime$). (b) Autocorrelations at several lags ($AC(\tau)$) as obtained from $p^\prime$ for the moving window of size 0.6 seconds. Different colors indicate $AC(\tau)$ having different time stamps. (c) time series of $var(AC)$ computed for different lags, equivalent to integer multiples of the acoustic time period. }
    \label{fig:varAC_convergence}
\end{figure}

For a given time series $p(t)$, the coefficient of lag-$\tau$ autocorrelation is defined as:
\begin{equation}
    AC(\tau)=\frac{\sum_{i=1}^{N-k}(p(i)-\Bar{p})(p(i+\tau)-\Bar{p})}{\sum_{i=1}^{N}(p(i)-\Bar{p})^2},
\end{equation}
where $\Bar{p}=\sum_{i=1}^{N}p(i)$. The variance of autocorrelations is defined as:
\begin{equation}
    var(AC)=\frac{1}{N-1} \sum_{\tau=1}^{k}(AC(\tau)-\Bar{AC})^2,
\end{equation}
where $\Bar{AC}=\sum_{\tau=1}^{k}AC(\tau)$. The value of $var(AC)$ depends on the number of $lag-\tau$ since time series separated by significantly large lags are less correlated to each other, as evident from figure~\ref{fig:varAC_convergence}(b). 
In order to choose the ideal number of lags, we compute $var(AC)$ for lags equivalent to integer multiples of the acoustic cycle. Each cycle contains approximately 45 samples. In figure~\ref{fig:varAC_convergence}(b), we compute var(AC) for lags equivalent to 2 to 10 acoustic cycles for a prototypical non-autonomous acoustic pressure fluctuation time series. For a perfectly periodic signal, var(AC) would be 0.5. Therefore, we should get $var(AC)\approx 0.5$ at the state of thermoacoustic instability. From figure~\ref{fig:varAC_convergence}(c), it is evident that $var(AC)$ converges to $0.5$ for two or fewer acoustic cycles (indicated by blue and cyan dot-dashed lines). Hence, we select $k=90$ number of lags (equivalent to 2 acoustic cycles since each cycle contains 45 sample points) for computing $var(AC)$.

\begin{figure}[ht]
    \centering
    \includegraphics[height=!,width=0.5\textwidth]{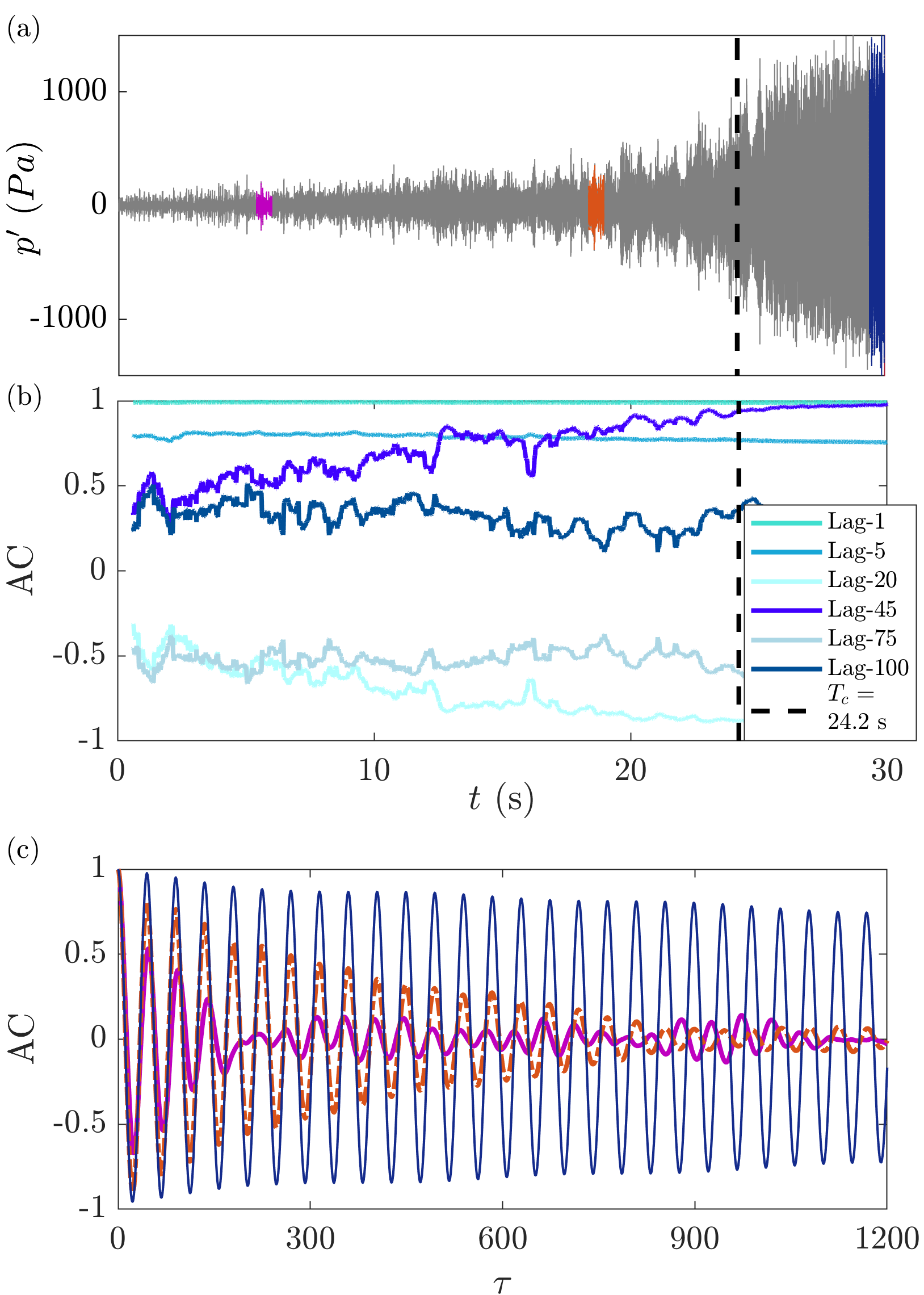}
    \caption{(a) A prototypical time series for the rate of change of Reynolds number $r=1542$ s$^{-1}$. Parts of the time series colored in pink, orange, and blue correspond to the states of aperiodic chaotic oscillations, intermittency, and periodic oscillations. The dashed black line indicates the critical time $(T_c)=24.2$ s for transitions to the state of periodic oscillations. (b) Time series of autocorrelations at different lags. (c) Autocorrelations for different lags corresponding to different dynamical states observed in experiments.}
    \label{fig:supplementary_fig3}
\end{figure}
\bibliographystyle{apsrev4-2} 
\bibliography{bibliography_file_EWS}